\documentclass{lmcs}
\pdfoutput=1

\usepackage{lastpage}
\lmcsdoi{15}{3}{30}
\lmcsheading{}{\pageref{LastPage}}{}{}%
{Jul.~04,~2017}{Sep.~20,~2019}{}


\usepackage{amssymb}
\usepackage{mathpartir}

\newcommand{\wpop}{\mathop{\wp}}
\newcommand{\vwpop}{\mathop{\vec\wp}}
\newcommand{\votimes}{\mathop{\vec\otimes}}

\usetikzlibrary{positioning}
\usetikzlibrary{calc}

\begin{document}

\title{On noncommutative extensions of linear logic}
\author{Sergey Slavnov}
\address{National Research University Higher School of Economics}
\email{sslavnov@yandex.ru}

\begin{abstract}
\emph{Pomset logic} introduced by Retor\'e is an extension of linear logic with a self-dual
noncommutative connective. The logic is defined by means of proof-nets, rather than a sequent calculus.
Later a deep inference system
 BV  was developed with an eye to capturing Pomset logic,
 but equivalence of system has not been proven up to now. As for a
 sequent calculus formulation, it has not been known for either of these logics, and there are convincing arguments that
 such a sequent calculus  in the usual sense simply does not exist for them.

In an on-going work on semantics we discovered a system similar to Pomset logic, where a  noncommutative connective is
no
 longer self-dual. Pomset logic appears as a degeneration, when the class of models is restricted.

Motivated by these semantic considerations, we define in the current work  a \emph{semicommutative multiplicative linear
 logic}, which is multiplicative linear logic extended with two nonisomorphic noncommutative connectives (not to be
 confused with very different Abrusci-Ruet noncommutative logic). We develop a syntax of
  proof-nets and show how this logic degenerates to Pomset logic.

However, a more interesting problem than just finding yet another noncommutative logic is to find  a sequent calculus for
 this logic. We introduce \emph{decorated sequents}, which are sequents equipped with an extra structure of a binary
 relation of \emph{reachability} on formulas. We define a decorated sequent calculus for semicommutative logic and prove
 that it is cut-free, sound and complete. This is adapted to ``degenerate'' variations, including Pomset logic. Thus, in
 particular, we give a variant of sequent calculus formulation for Pomset logic, which is one of the key results of the
 paper.
\end{abstract}

\maketitle

\section{Introduction}
There is a number of noncommutative variations and extensions of linear logic known in literature,
starting from Lambek
 calculus~\cite{Lambek}, which historically precedes linear logic itself. Probably the best known and best behaved
 is \emph{cyclic linear logic} of Yetter~\cite{Yetter}. Let us mention also \emph{noncommutative logic} of Abrusci and
  Ruet~\cite{Abrusci, AbrusciRuet}, which mixes cyclic logic and the ordinary, commutative linear logic.
  Another approach
  to combining commutative and noncommutative features is based on introducing \emph{exchange modalities} in Lambek calculus~\cite{JiangEadesPaiva,EadesPaiva}.

The topic of the current paper, however, is somewhat far from the above systems. Rather, it is related to linear logic noncommutative extensions stemming from the seminal work~\cite{Retore} of Retor\'e. Retor\'e found an extension of linear logic (technically speaking, of linear logic with the Mix rule) with a self-dual binary associative noncommutative connective \emph{seq} or \emph{before}, denoted as $<$, ``intermediate'' between times and par, so that in the corresponding logic it holds that
\begin{equation}\label{seq}
A\otimes B\vdash A<B\vdash A \wpop B.
\end{equation}
(It seems though that there is no firm consensus  on a standard notation for this connective; different works~\cite{Retore,BluteSlavnovPanangaden,GuglielmiStrassburger} use different symbols.)

The system found by Retor\'e, \emph{Pomset logic} was based on semantics (it has a denotational model in the category of \emph{coherent spaces}) and is defined by means of proof-nets. However a traditional Gentzen-style sequent calculus formulation was missing. Later, a \emph{deep inference} system \emph{BV} was designed~\cite{Guglielmi} to capture Pomset logic. Unfortunately, equivalence of systems was not proven and this still remains an open problem, although it is known that BV is contained in Pomset logic~\cite{StrassburgerThesis}. (In fact, rules of BV directly translate into a system of graph rewrite rules which transform proof-nets to proof-nets. This rewriting system itself was proposed in~\cite{Retore99}.)

As for a sequent calculus formulation, in~\cite{Tiu}, it was proved that, in fact, no sequent calculus in the usual sense can capture these logics.

 The system BV was subsequently extended~\cite{GuglielmiStrassburger} to accommodate linear logic exponentials,  an extension with additive connectives was considered as well~\cite{Horne}. Basically, BV together with other deep inference systems gave rise to a rather active
 research field\footnote{Consult {http://alessio.guglielmi.name/res/cos/ for the current state}.}.

\subsection{Interpretation in probabilistic coherence spaces}
Pomset logic was introduced in~\cite{Retore} together with a denotational model in the category of \emph{coherent spaces},
 which satisfies quite strong completeness properties~\cite{Retore94}.
Since the system BV is contained in Pomset logic, it follows that BV can be modeled in coherent spaces as well.
In~\cite{BluteSlavnovPanangaden} an attempt was made to give a general category-theoretic axiomatization of BV, and an abstract notions of a \emph{BV category} was introduced  (although no kind of soundness result was proven).

In particular, a concrete example of a BV category was constructed in the setting of \emph{probabilistic coherence spaces} (PCS).

PCS were first introduced in~\cite{Girard_Logic_quantic} as a model for the multiplicative-additive linear logic, and later much studied in~\cite{DanosErhahrdPCS}, where the model was extended to encompass exponentials.
%In~\cite{BluteSlavnovPanangaden} the following setting was used.
PCS can be seen as a ``weighted'' version of ordinary coherent spaces, where relations are replaced with nonnegative real-valued functions, and intersection is replaced with \emph{pairing} of functions
 \[\langle f,g\rangle=\sum f(x)g(x).\]

Thus, a \emph{probabilistic coherence space} $A$ is a pair $A=(X,A)$, where $X$, the \emph{carrier}, is a  set, and $A\subseteq{\bf R}_+^X$ is a set of functions from $X$ to ${\bf R}_+$ coinciding with its
%\begin{itemize}
%  \item for any $v\in{\bf R}_+^X$ there exist $\lambda,\mu>0$ such that $\lambda v\in A$, $\mu v\not\in A$;
%  \item the set $A$ coincides with its
\emph{bipolar} $A^{\bot\bot}$, where the \emph{polar} $S^\bot$ of any set $S\in{\bf R}_+^X$ is defined as
  \begin{equation}\label{polar}
  S^{\bot}=\{v\in{\bf R}_+^X|\mbox{ }\forall u\in S\mbox{ }\langle v,u\rangle\leq 1\}.
  \end{equation}
  %the pairing $\langle f,g\rangle$ being \[\langle f,g\rangle=\sum\limits_{x\in X}f(x)g(x).\]
%\end{itemize}
(In the setting of~\cite{BluteSlavnovPanangaden}, the set $X$ was assumed to be finite.)
%(This is precisely the definition currently used~\cite{DanosErhahrdPCS}, specialized to the case of finite $X$.)
 %was built, as an addition to Retor\'e's original model based on ordinary coherent spaces.

 %Multiplicative connectives are readily interpreted in PCS.

 Then linear negation is interpreted in PCS by means of (\ref{polar}).
 Further, if \[A=(X,A),\quad B=(Y,B)\] are PCS, then the spaces $A\multimap B$, $A\otimes B$, $A\wpop B$, all with the same carrier  $X\times Y$, are  defined by
 \[A\multimap B= \Big\{F\in {\bf R}_+^{X\times Y}|\mbox{ }\forall u\in A\sum\limits_{x\in X}F(x,y)(x)u(x)\in B \Big\},\]
 \[A\wpop B=A^\bot\multimap B,\quad A\otimes B={(A^\bot\wpop B^\bot)}^\bot.\]

 Elements of the set $A\multimap B\subseteq{\bf R}_+^{X\times Y}$ become morphisms from $A$ to $B$, which are composed by means of the formula
%
%  For $A=(X,A)$, $B=(Y,A)$, $C=(Z,C)$, the composition of $F\in A\multimap B$, $G\in B\multimap C$ is given by
 \[G\circ F(x,z)=\sum\limits_{y\in Y}F(x,y)G(y,z).\]

 It turns out that there is also an associative noncommutative self-dual  \emph{seq} operation (denoted in~\cite{BluteSlavnovPanangaden} as $\oslash$) on PCS, at least in the finite-dimensional case.

 If $A=(X,A)$, $B=(Y,B)$ are PCS then the PCS $A\oslash B$ with the carrier $X\times Y$ is defined by
 \begin{equation}\label{seq in PCS}
 A\oslash B=\Big\{\sum\limits_{i\in I} u_i\otimes v_i|\mbox{ }\forall i\in I\mbox{ } u_i\in A\mbox{ and }\sum\limits_{i\in I}v_i\in B\Big\}.
 \end{equation}
 (In the above formula, $I$ is an arbitrary finite index set, and
  we use the identification ${\bf R}^{X\times Y}\cong{\bf R}^{X}\otimes{\bf R}^{Y}$.)

  It was shown that the seq operation satisfies basic properties of the seq connective in BV and makes  the category of PCS  a BV category.

  \subsection{PCS and partially ordered vector spaces}
  On the other hand, as it was articulated, for example, in~\cite{DanosErhahrdPCS}, probabilistic coherence spaces can be seen as a special case of \emph{partially ordered vector spaces} (POVS).

  Indeed, for any set $X$, the set of functions ${\bf R}^X$ is partially ordered by
  \[f\leq g\mbox{ iff }\forall x\in X f(x)\leq g(x),\]
  and the subset ${\bf R}^X_+$ is the set of nonnegative elements, the \emph{nonnegative cone}.

  Then, a PCS $A=(X,A)$ can be described as a partially ordered vector space $V$ of a specific form
  \begin{equation}\label{PCS as POVS}
  V={\bf R}^X
  \end{equation}
  together with a subset $A$ of its nonnegative cone.

 Remarkably, it can be shown that in the finite-dimensional case the interpretation of multiplicative linear logic extends from PCS to general POVS more or less verbatim.

 In an on-going work the author of this paper discovered that the above seq operation defined by (\ref{seq in PCS}) can also be extended to general finite-dimensional POVS\@. This will be shown in a forthcoming paper. However,  this operation is no longer self-dual, and thus gives rise to two non-isomorphic (noncommutative associative) dual operations, which it seems reasonable to denote as $\votimes$ and $\vwpop$.

 The two connectives degenerate into one, when condition (\ref{PCS as POVS}) is imposed on every  object $V$.

 \subsection{Semicommutative logic}
 The above discussion of semantics  motivates out interest in constructing a logic with two dual noncommutative connectives, such that Pomset logic can be seen as its degenerate variant.

 We call the system we propose \emph{semicommutative multiplicative linear logic}.
  The principle (\ref{seq}) changes to
  \[{(A\votimes B)}^\bot=A^\bot\vwpop B^\bot,\quad {(A\vwpop B)}^\bot=A^\bot\votimes B^\bot,\]
\begin{equation}\label{n.c.seq}
A\otimes B\vdash A\votimes B,\mbox{ }A\vwpop B\vdash A\wpop B,
\end{equation}
or, in presence of the Mix rule, to
\begin{equation}\label{n.c.seq_with_mix}
A\otimes B\vdash A\votimes B\vdash A\vwpop B\vdash A\wpop B.
\end{equation}
In this  paper we define semicommutative logic by means of  proof-nets and show how it ``degenerates'' into
 Pomset logic by first adding the Mix rule and then declaring the two noncommutative connectives isomorphic.

\subsection{Decorated sequents}
However, a more interesting problem than just finding yet another noncommutative logic is to find a sequent calculus for this logic. This is nontrivial; for example, it is proven that  system BV  does not admit  sequent calculus formulation~\cite{Tiu} at all.  This result apparently applies as well to Pomset logic, which conjecturally is the same as BV (this is discussed in~\cite{Pogodalla_apomset}). And  semicommutative logic of this paper is closely related to these systems.

 Thus it seems unreasonable to look for a formulation of semicommutative logic as a sequent calculus in the ordinary sense.

 We consider \emph{decorated} sequents, which are sequents equipped with an extra structure of a binary relation of \emph{reachability} on formulas. We define a decorated sequent calculus for semicommutative logic and prove that it is cut-free, sound and complete. This is adapted to ``degenerate'' variations, including Pomset logic. Thus, in particular, we give a (sort of) sequent calculus formulation for Pomset logic, and this is one of the key results of the paper. (We say ``sort of'' sequent calculus, because decorated sequents are not sequents in the usual sense of the word.)

To conclude the introduction, we should remark on an important difference between semicommutative logic of this paper
and some other   noncommutative   extensions of linear logic.
Concretely, we want to avoid confusion with  the above-mentioned system of  Abrusci-Ruet~\cite{Abrusci,AbrusciRuet}.

Formally, both  systems have two pairs of multiplicative connectives, one commutative and one noncommutative. This may create a suspicion of some relation, which would be strange because Abrusci-Ruet noncommutative logic is a very complex and nontrivial construction compared with the semicommutative logic of this paper. However, a formal resemblance between the two systems is deceptive. In noncommutative logic the noncommutative pair of connectives is almost as ``powerful'' as the commutative one; both commutative and noncommutative par do define (different) linear implications. (On the  sematic side: each pair corresponds to a separate $*$-autonomous structure on the modeling category, see~\cite{BluteLamarcheRuet}). In semicommutative logic we have nothing of this kind. The noncommutative pair is just an extra bimonoidal structure with no particular power.

\section{Semicommutative linear logic}
We assume that the reader is familiar with multiplicative linear logic (MLL) (see~\cite{Girard} for an introduction), including the variation with the Mix rule (MLL+Mix)~\cite{FR}, as well as with proof-nets and the Danos-Regnier criterion~\cite{DR}, both for MLL and MLL+Mix~\cite{FR}.

In this  section we introduce \emph{semicommutative} linear logic, a simple extension of MLL with two noncommutative connectives.

\subsection{Language}
The \emph{language} of \emph{semicommutative linear logic} (ScMLL) is that of MLL supplied with two new binary connectives $\votimes$, $\vwpop$.

More accurately, we assume that we are given  a set $N$ of  \emph{positive literals}. We define then the set $N^\bot$ of \emph{negative literals}
as
\[N^\bot=\{X^\bot|\mbox{ }X\in N\}.\]
Elements of $N\cup N^\bot$ are  called literals.

Formulas of ScMLL are defined by the following induction.
\begin{itemize}
  \item Any $X\in N\cup N^\bot$ is a formula;
  \item if $X$, $Y$ are   formulas, then $X\wpop Y$, $X\otimes Y$, $X\vwpop Y$ and $X\votimes Y$ are   formulas;
  \end{itemize}

\subsection{Proof-nets}
We define semicommutative linear logic by means of proof-nets. As usual, we define proof-nets in two steps. First we define proof-structures.

Recall that a \emph{mixed graph} $G$ is a triple $G=(V,E,A)$, where $V$ is a set of \emph{vertices}, $E$, the set of
\emph{undirected edges}, is a set of two element subsets of $V$, and $A\subseteq V\times V$, the set of
 \emph{directed edges}, is a set of ordered pairs of distinct elements of $V$.
In plainer, but less accurate language, a mixed graph is a graph that may contain both directed and undirected edges.

A \emph{proof-structure} $\rho$ is a mixed graph,   whose vertices are labeled by ScMLL formulas, and some of whose vertices are selected as \emph{conclusions}, built inductively by the following rules.

A graph with two vertices labeled by dual propositional symbols $p$ and $p^\bot$ connected by one undirected edge is a proof-structure with conclusions $p$, $p^\bot$.

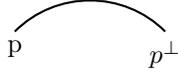
\[
\begin{tikzpicture}
\node[below,inner sep=2mm] (p) {$\strut p$};
\node[below,inner sep=2mm,right=of p] (pbot) {$\strut p^\bot$};
\draw [thick] ($(p.north) + (0,-0.2)$) edge[bend left] ($(pbot.north) + (-0.15,-0.2)$);
\end{tikzpicture}
\]

Such a proof-structure is called an \emph{axiom link}.

A disjoint union of two proof-structures $\rho_1$ and $\rho_2$ is a proof-structure whose conclusions are those of $\rho_1$ or $\rho_2$.

If $\rho$ is a proof-structure whose set of conclusions contains two (vertices labeled by) formulas $A$ and $B$, then the labeled graph $\rho'$  obtained from $\rho$ by attaching to $A$ and $B$  one of the following labeled graphs,

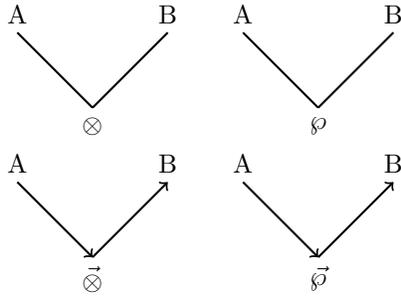
\begin{mathpar}
\begin{tikzpicture}
\node[above] at (0,0) {$A$};
\node[above] at (2,0) {$B$};
\draw [thick] (0,0) --(1,-1);
\draw [thick] (2,0) --(1,-1);
\node[below] at (1,-1) {$\otimes$};
\end{tikzpicture}
\and
\begin{tikzpicture}
\node[above] at (0,0) {$A$};
\node[above] at (2,0) {$B$};
\draw[thick](0,0) --(1,-1);
\draw[thick] (2,0) --(1,-1);
\node[below] at (1,-1) {$\wpop$};
\end{tikzpicture}
\and
\begin{tikzpicture}
\node[above] at (0,0) {$A$};
\node[above] at (2,0) {$B$};
\draw[thick, ->] (0,0) --(1,-1);
\draw [thick, ->](1,-1) --(2,0);
\node[below] at (1,-1) {$\votimes$};
\end{tikzpicture}
\and
\begin{tikzpicture}
\node[above] at (0,0) {$A$};
\node[above] at (2,0) {$B$};
\draw[thick,->] (0,0) --(1,-1);
\draw [thick,<-](2,0) --(1,-1);
\node[below] at (1,-1) {$\vwpop$};
\end{tikzpicture}
\end{mathpar}

called, respectively, an $\otimes$-link, a $\wpop$-link, a $\votimes$-link, a $\vwpop$-link, is a proof-structure. The new proof-structure $\rho'$ has the same conclusions as $\rho$ except for $A$ and $B$, and one new conclusion, respectively $A\otimes B$, $A\wpop B$, $A\votimes B$, $A\vwpop B$, depending on the link attached.

If $\rho$ is a proof-structure whose conclusions contain two (vertices labeled by) dual formulas $A$ and $A^\bot$, then the labeled graph $\rho'$  obtained from $\rho$ by attaching to $A$ and $A^\bot$  the following graph, called a \emph{Cut-link},
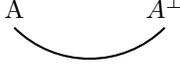
\[
\begin{tikzpicture}
\node (A) {A};
\node[right=of A] (Abot) {$A^\bot$};
\draw [thick] (A.south) edge[bend right] ($(Abot.south) + (-0.15,0)$);
%\draw [thick] (0,0) --(1,-1);
%\draw [thick] (2,0) --(1,-1);
%\node[below] at (1,-1) {Cut};
\end{tikzpicture}
\]
is a proof-structure.
The new proof-structure $\rho'$ has the same conclusions as $\rho$ except for $A$ and $A^\bot$.

Each $\otimes$-, $\wpop$-, $\votimes$- or $\vwpop$-link has three vertices, one of which is called the \emph{conclusion} of the link, and the two other, \emph{premises}. For example for $A\otimes B$, the vertices labeled by $A$ and $B$ are premises, and the remaining one, the conclusion, similarly for other types. Also, observe that, in $\otimes$- and $\wpop$-links, the edges forming the link are undirected.

A proof-structure without cut-links is called \emph{cut-free}.

We use the following terminology below. A \emph{generalized} $\wpop$-link is a $\wpop$-link or a $\vwpop$-link, a \emph{generalized} $\otimes$-link is a $\otimes$-link or a $\votimes$-link, an \emph{ordered} link is a $\vwpop$-link or a $\votimes$-link.

In order to define proof-nets we introduce more terminology.
Recall that an \emph{elementary path} in a graph is a path traversing each of its vertices exactly once.
An \emph{elementary cycle} is a cycle traversing exactly once each of its vertices except the starting and the ending one.

Now let $\rho$ be a proof-structure, and $\phi$ be an elementary  path, respectively, elementary cycle, in  $\rho$.
 A \emph{critical link} on $\phi$ is an ordered link such that $\phi$ goes through its conclusion and both premises.

  The path, respectively, cycle, $\phi$ is \emph{essentially directed}, if it goes through all its critical links in the direction of
 arrows. (For example for a $\vwpop$-link $A\vwpop B$, the path $\phi$ should go as $A$---$A\vwpop B$---$B$).

 We now proceed to defining proof-nets. We use switchings in the style of Danos-Regnier criterion.
 A \emph{switching} $\sigma$ of the proof-structure $\rho$ is a choice, for each generalized $\wpop$-link, of
 one of the two edges forming the link.
  The \emph{first round} of the switching $\sigma$ of the proof-structure $\rho$ is the undirected graph obtained
  from $\rho$ by erasing from each generalized $\wpop$-link the edge chosen by $\sigma$ and forgetting directions
   of remaining edges. The \emph{second round} of $\sigma$ is the directed graph obtained from $\rho$
   by erasing, for each $\wpop$-link the edge chosen by $\sigma$ (but not touching $\vwpop$-links, and keeping
   directions of the edges).

% To fix terminology: in this paper, a \emph{cycle} (in a graph) is a path whose endpoint vertices coincide, and which traverses all its other vertices exactly once.

\begin{defi}
    A proof-net is a proof-structure $\rho$, such that for any switching $\sigma$ the first round is connected
   and acyclic, and the second round has no essentially directed cycles.
\end{defi}

   A \emph{sequent} $\Gamma$ is a multiset of ScMLL formulas.

   \begin{defi} A sequent $\Gamma$ is derivable in ScMLL if there exists
   a proof-net $\rho$ whose multiset of conclusions is $\Gamma$. We say that $\rho$ is a {\bf proof-net of} (the sequent) $\Gamma$.
   A formula $A$ is derivable in ScMLL if there exists a proof-net whose only conclusion  is $A$.
   \end{defi}

When the sequent $\Gamma$ is derivable, we write
\[\vdash\Gamma.\]
We will also use a two-sided notation \[\Gamma\vdash\Delta\] meaning that the sequent $\Gamma^\bot,\Delta$ is derivable. Here, if \[\Gamma=A_1,\ldots,A_n\] then the sequent $\Gamma^\bot$ is defined as
\[\Gamma^\bot=A_1^\bot,\ldots,A_n^\bot.\]

\subsubsection{Examples}
We give some examples and non-examples of proof-nets.

The following proof-structure is not a proof-net.
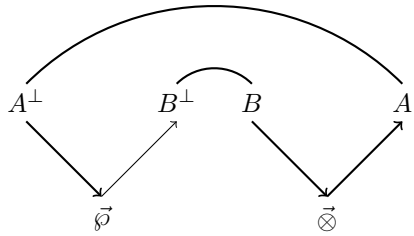
\[
\begin{tikzpicture}{scale =2}
\node[above] at (0,0) {$A^\bot$};
\node[above] at (2,0) {$B^\bot$};
\draw[thick,->] (0,0) --(1,-1);
\draw[thick,<-](2,0) --(1,-1);
\node[below] at (4,-1) {$\votimes$};
\node[above] at (3,0) {$B$};
\node[above] at (5,0) {$A$};
\draw[thick,->] (3,0) --(4,-1);
\draw [thick,<-](5,0) --(4,-1);
\node[below] at (1,-1) {$\vwpop$};
\draw [thick](0,0.5) to [out=45,in=135] (5,0.5);
%\node[above] at (2.5,1){Id};
\draw [thick](2,0.5) to [out=45,in=135] (3,0.5);
\end{tikzpicture}
\]

The first round of any of the two possible switchings is indeed acyclic and connected. However, in the second round no edge is erased and we get a directed cycle.
\smallskip

The following is a proof-net.
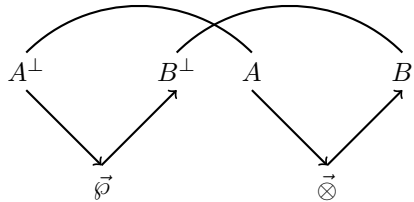
\[
\begin{tikzpicture}{scale =2}
\node[above] at (0,0) {$A^\bot$};
\node[above] at (2,0) {$B^\bot$};
\draw[thick,->] (0,0) --(1,-1);
\draw [thick,<-](2,0) --(1,-1);
\node[below] at (1,-1) {$\vwpop$};
\node[above] at (3,0) {$A$};
\node[above] at (5,0) {$B$};
\draw[thick,->] (3,0) --(4,-1);
\draw [thick,<-](5,0) --(4,-1);
\node[below] at (4,-1) {$\votimes$};
\draw [thick](0,0.5) to [out=45,in=135] (3,0.5);
%\node[above] at (2.5,1){Id};
\draw [thick](2,0.5) to [out=45,in=135] (5,0.5);
\end{tikzpicture}
\]
There is a cycle in the second round of any switching, but it is not essentially directed.

\smallskip
The following is not a proof-net.
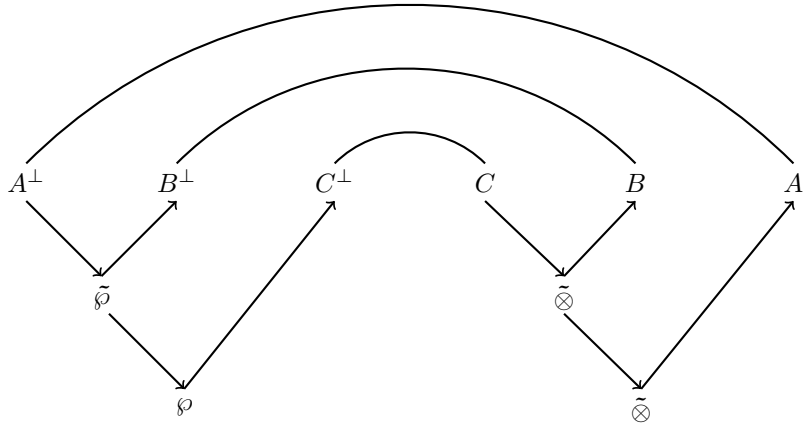
\[
\begin{tikzpicture}{scale=2}
\node[above] at (0,0) {$A^\bot$};
\node[above] at (2,0) {$B^\bot$};
\node[above] at (4.1,0) {$C^\bot$};
\draw[thick, ->] (0,0) --(1,-1);
\draw [thick,->] (1,-1)--(2,0);
\node[below] at (1,-1) {${\bf \vwpop}$};
\draw[thick,->] (1.1,-1.5) --(2.1,-2.5);
\draw[thick,->] (2.1,-2.5)--(4.1,0);
\node[below] at (2.1,-2.5) {${\bf \wpop}$};
\node[above] at (6.1,0) {$C$};
\node[above] at (8.1,0) {$B$};
\node[above] at (10.2,0) {$A$};
\draw[thick,->] (6.1,0) --(7.15,-1);
\draw[thick,->] (7.15,-1)--(8.1,0);
\node[below] at (7.15,-1) {${\bf \votimes}$};
\draw[thick,->] (7.15,-1.5) --(8.175,-2.5);
\draw[thick,->] (8.175,-2.5)--(10.2,0);
\node[below] at (8.175,-2.5) {${\bf \votimes}$};
\draw[thick] (0,0.5) to [out=45,in=135] (10.2,0.5);
%\node[above] at (2.5,1){Id};
\draw[thick] (2,0.5) to [out=45,in=135] (8.1,0.5);
\draw[thick] (4.1,0.5) to [out=45,in=135] (6.1,0.5);
\end{tikzpicture}
\]
It passes the first round, however in the second round of any switching we get a cycle (shown dashed below). Observe that the cycle is not directed, but it is essentially directed.
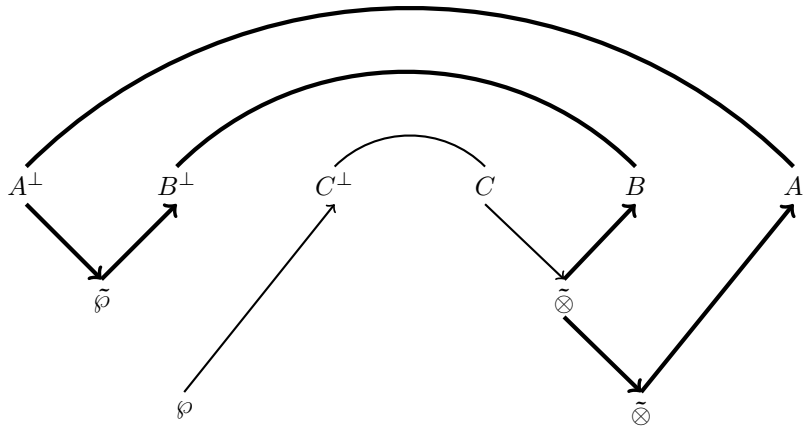
\[
\begin{tikzpicture}{scale=2}
\node[above] at (0,0) {$A^\bot$};
\node[above] at (2,0) {$B^\bot$};
\node[above] at (4.1,0) {$C^\bot$};
\draw [thick,dashed, ->] (0,0) --(1,-1);
\draw [thick,dashed,->] (1,-1)--(2,0);
\node[below] at (1,-1) {${\bf \vwpop}$};
%\draw[thick,->] (1.1,-1.5) --(2.1,-2.5);
\draw[thick,->] (2.1,-2.5)--(4.1,0);
\node[below] at (2.1,-2.5) {${\bf \wpop}$};
\node[above] at (6.1,0) {$C$};
\node[above] at (8.1,0) {$B$};
\node[above] at (10.2,0) {$A$};
\draw[thick,->] (6.1,0) --(7.15,-1);
\draw[thick,dashed,->] (7.15,-1)--(8.1,0);
\node[below] at (7.15,-1) {${\bf \votimes}$};
\draw[thick,dashed,->] (7.3,-1.55) --(8.175,-2.5);
\draw[thick,dashed,->] (8.175,-2.5)--(10.2,0);
\node[below] at (8.175,-2.5) {${\bf \votimes}$};
\draw[thick,dashed] (0,0.5) to [out=45,in=135] (10.2,0.5);
%\node[above] at (2.5,1){Id};
\draw[thick,dashed] (2,0.5) to [out=45,in=135] (8.1,0.5);
\draw[thick] (4.1,0.5) to [out=45,in=135] (6.1,0.5);
\end{tikzpicture}
\smallskip
\]
The following is a proof-net, as can be found by inspection.
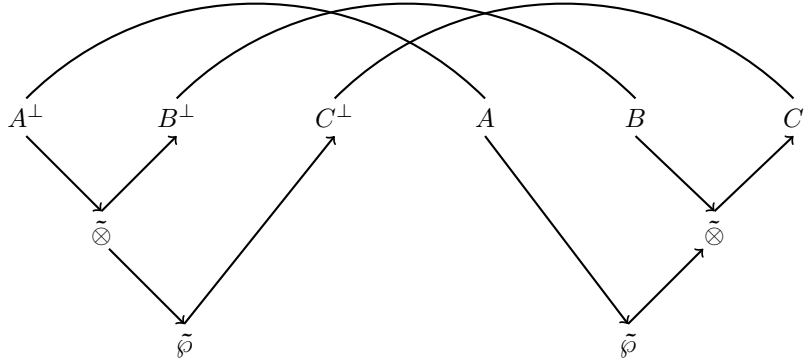
\[
\begin{tikzpicture}{scale=2}
\node[above] at (0,0) {$A^\bot$};
\node[above] at (2,0) {$B^\bot$};
\node[above] at (4.1,0) {$C^\bot$};
\draw[thick, ->] (0,0) --(1,-1);
\draw [thick,->] (1,-1)--(2,0);
\node[below] at (1,-1) {${\bf \votimes}$};
\draw[thick,->] (1.1,-1.5) --(2.1,-2.5);
\draw[thick,->] (2.1,-2.5)--(4.1,0);
\node[below] at (2.1,-2.5) {${\bf \vwpop}$};
\node[above] at (6.1,0) {$A$};
\node[above] at (8.1,0) {$B$};
\node[above] at (10.2,0) {$C$};
\draw[thick,->] (8.1,0) --(9.15,-1);
\draw[thick,->] (9.15,-1)--(10.2,0);
\node[below] at (9.15,-1) {${\bf \votimes}$};
\draw[thick,->] (8,-2.5) --(9,-1.5);
\draw[thick,->] (6.1,0)--(8,-2.5);
\node[below] at (8,-2.5) {${\bf \vwpop}$};
\draw[thick] (0,0.5) to [out=45,in=135] (6.1,0.5);
%\node[above] at (2.5,1){Id};
\draw[thick] (2,0.5) to [out=45,in=135] (8.1,0.5);
\draw[thick] (4.1,0.5) to [out=45,in=135] (10.2,0.5);
\end{tikzpicture}
\]

\smallskip
   We collect some observations on principles derivable and non-derivable in ScMLL\@.

   \begin{rem}\label{what_is_provable} The logic ScMLL contains MLL\@. The connectives $\vwpop$ and $\votimes$ in ScMLL are provably associative.
   ScMLL derives (\ref{n.c.seq}) and the following principles
      \[A\otimes (B\vwpop C )\vdash (A\otimes B)\vwpop C, \mbox{ }
   (A\wpop B)\votimes  C\vdash A\wpop (B\votimes  C),\]
\[A\votimes (B\wpop C )\vdash (A\votimes B)\wpop C, \mbox{ }
   (A\vwpop B)\otimes  C\vdash A\vwpop (B\otimes  C),\]
\[A\votimes (B\vwpop C )\vdash (A\votimes B)\vwpop C, \mbox{ }
   (A\vwpop B)\votimes  C\vdash A\vwpop (B\votimes  C).\]
   ScMLL does not derive $A\votimes B\vdash B\votimes A$. \qed%
   \end{rem}

\subsubsection{Commutative projections}
Observe that the first round part of the above proof-net definition  is just the  Danos-Regnier criterion~\cite{DR} for ordinary
   MLL proof-nets. In particular, it follows that, if in the proof-net $\rho$ we forget  direction of edges and erase all
   arrows above connectives,
   we get a correct proof-net $\rho'$ for an ordinary MLL\@.

   Given a proof-net $\rho$  we define  the \emph{commutative projection}  $\rho'$ of $\rho$ as the proof-structure obtained from $\rho$ by making all edges undirected and replacing all $\votimes$ links with $\otimes$ links, and all  $\vwpop$ links with $\wpop$ links.

It is easy to see that thus obtained $\rho'$ is a proof-net.

   \subsection{Cut-elimination}
   \begin{lem}\label{ScMLL cut-reduction} There is an algorithm transforming any proof-net with cut-links to a cut-free proof-net with the same
   conclusions.
   \end{lem}

    \begin{proof} The algorithm is, of course, that of ordinary MLL as far as unordered links are concerned.
   Thus if a cut-link is between dual propositional symbols $p$, $p^\bot$, then, by the acyclicity condition for
   switchings in the first round, it necessarily has the form
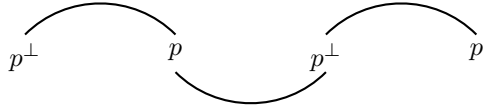
\[
\begin{tikzpicture}{}
\node (pbot0) at (0,0) {$\strut p^\bot$};
\node[right=of pbot0] (p0) {$\strut p$};
\node[right=of p0] (pbot1) {$\strut p^\bot$};
\node[right=of pbot1] (p1) {$\strut p$};
\draw [thick] (pbot0.north) edge[bend left] ($(p0.north) + (0,-0.15)$);
\draw [thick] ($(p0.south) + (0, 0.1)$) edge[bend right] ($(pbot1.south) + (-0.15, 0.1)$);
\draw [thick] (pbot1.north) edge[bend left] ($(p1.north) + (0,-0.15)$);
\end{tikzpicture}
\]
   and can be replaced by an axiom link.
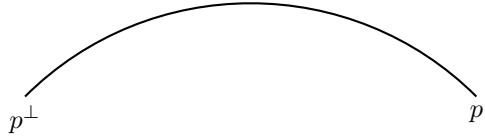
\[
\begin{tikzpicture}{}
\node (pbot) {$\strut p^\bot$};
\node[right=of pbot] (p) {$\strut p$};
\draw[thick] (pbot.north) edge[bend left] (p.north);
\end{tikzpicture}
\]
   If a cut-link is between two compound formulas, it is replaced by two cut-links between their subformulas.

   The generic case is when the compound formulas are of the form $A\votimes B$,
   $A^\bot\vwpop B^\bot$.
 So assume that we have a proof-net $\rho$ with a cut-link as shown below.
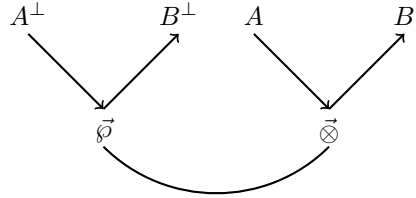
\[
\begin{tikzpicture}
\node[above] at (0,0) {$A^\bot$};
\node[above] at (2,0) {$B^\bot$};
\draw[thick,->](0,0)--(1,-1);
\draw[thick,->](1,-1)--(2,0);
\node[below] at (1,-1){$\vwpop$};
\node[above] at (3,0) {$A$};
\node[above] at (5,0) {$B$};
\draw[thick,->](3,0)--(4,-1);
\draw[thick,->](4,-1)--(5,0);
\node[below] at (4,-1){$\votimes$};
\draw [thick] (1,-1.6) to [out=-45,in=-135] (4,-1.6);
%\draw[thick](1,-1.5)--(2.5,-2.5)--(4,-1.5);
%\node[below] at (2.5,-2.5){Cut};
\end{tikzpicture}
\]
Let us denote the above subgraph of $\rho$ as $s$.
   The subgraph $s$ is  replaced by a new subgraph $s'$ consisting of two cut-links, as shown below.
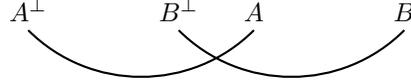
\[
\begin{tikzpicture}
\node[above] at (0,0) {$A^\bot$};
\node[above] at (2,0) {$B^\bot$};
\draw [thick] (0,0) to [out=-45,in=-135] (3,0);
\draw [thick] (2,0) to [out=-45,in=-135] (5,0);
%\draw[thick](0,0)--(1.5,-1)--(3,0);
%\draw[thick](2,0)--(3.5,-1)--(5,0);
%\node[below] at (1.5,-1){Cut};
\node[above] at (3,0) {$A$};
\node[above] at (5,0) {$B$};
%\node[below] at (3.,-1){Cut};
\end{tikzpicture}
\]
   This gives us a new proof-structure $\rho'$ and we need to check that $\rho'$ is a proof-net.

   %Thus we need to check
%   that, if $\rho$ is the original proof-net with a cut-link between compound formulas as above, then the proof-structure $\rho'$
%   obtained from $\rho$ by the above reduction is a proof-net.

   Now, the first-round part of proof-net definition   depends only
   on commutative projections. So $\rho'$ satisfies the first-round part if  $\rho$ does, because for their
   commutative projections this is a reduction of ordinary MLL proof-nets, and this takes proof-nets to proof-nets.
   So we only need to care about the second-round part.

   So assume that for some switching $\sigma$ of $\rho'$ there is an essentially directed cycle $\phi$ in the
   second round.
   Let us denote the common part of $\rho$ and $\rho'$ as $\tilde\rho$. I.e.
    $\tilde\rho$ is $\rho'$ without the cut-links between
   $A$ an $A^\bot$ and
   between $B$ and $B^\bot$.
   Let $\tilde\sigma$ be the switching of $\rho$, obtained by restricting $\sigma$
   to $\tilde\rho$ and extending this to  $\rho$ in an arbitrary way. (The extension consists in
   choosing one of the two edges in the $\vwpop$-link between $A^\bot$ and $B^\bot$, and it does not play role
   in the second round anyway.)

   The cycle $\phi$  must pass through $s'$, otherwise the cycle is present already in $\rho$ for the switching
   $\tilde\sigma$.
   Let us consider all possibilities.

   The cycle $\phi$ does not meet  $A$. Then $\phi$ consists of
   an essentially directed path $\phi_0$ in $\tilde\rho$ between $B$ and $B^\bot$ and the cut-link. Then $\phi_0$
   together with the path $B-A\votimes B-\mbox{Cut}-A^\bot\vwpop B^\bot-B^\bot$ create an essentially directed
   cycle for $\tilde\sigma$ in $\rho$ in the second round. So $\rho$ is not a proof-net.

   The cycle $\phi$ meets $A$ and enters $s'$ through $A$ (remember that $\phi$ has direction). Let $X$ be the vertex
   of $s'$ such that $\phi$ passes through $X$ before entering $A$ and does not go through any vertex of $s'$ between $X$ and $A$. Then there is an essentially
   directed path $\phi_0$ from $X$ to $A$ in $\tilde\rho$. There are three possibilities
   for $X$: $X$ can be $A^\bot$, $B$ or $B^\bot$. In all three possible cases the essentially directed
   $\phi_0$ is completed to an essentially directed cycle for $\tilde\sigma$ in the
   second round by attaching an essentially directed
   path from $A$ to $X$ in $s$. Again, it follows that $\rho$ is not a proof-net.

   The cycle meets $A$ and leaves $s'$ through $A$. Then $\phi$ enters $s'$ through $A^\bot$. Similarly to the
   preceding paragraph, let $X$ be the vertex of $s'$ passed by $\phi$  before $A^\bot$. Then there is an
   essentially directed path $\phi_0$ from $X$ to $A^\bot$ in $\tilde\rho$. Again we consider all three possibilities
   for $X$, and each of them gives rise to an essentially directed cycle for $\tilde\sigma$ in the
   second round. Hence $\rho$ is not a proof-net.

   It follows that such $\sigma$  and $\phi$ do not exist and $\rho'$ is a proof-net.
   The case of a cut-link between formulas $A\wpop B$ and $A\otimes B$ is treated similarly.
   \end{proof}
%   \begin{tikzpicture}{scale =2}
%\node[above] at (0,0) {$A^\bot$};
%\node[above] at (2,0) {$B^\bot$};
%\draw[thick](0,0)--(1,-1);
%\draw[thick](1,-1)--(2,0);
%\node[below] at (1,-1){$\vwpop$};
%\node[above] at (3,0) {$A$};
%\node[above] at (5,0) {$B$};
%\draw[thick](3,0)--(4,-1);
%\draw[thick](4,-1)--(5,0);
%\node[below] at (3,-1){$\votimes$};
%\draw[thick](1,-1.5)--(2,-2.5)--(3,-1.5);
%\end{tikzpicture}
   \smallskip

   \begin{cor}
   A sequent is derivable in ScMLL iff it is derivable by means of a cut-free proof-net.

   ScMLL is a conservative extension of MLL\@. \qed%
   \end{cor}

In the next section we give a sequent calculus formulation of ScMLL\@.

\section{Sequent calculus}
We want now to formulate ScMLL as a sequent calculus.

%There exists, however, a well-founded consensus that noncommutative multiplicative systems do not admit sequent calculus formulations  in the traditional sense with sequents being multisets or linearly ordered arrays of formulas. Basically, the only multiplicative binary connectives that we can introduce in such a setting are familiar MLL $\otimes$ and $\wpop$.

There are, however, strong arguments
that
an adequate formulation of semicommutative logic in the language of ordinary sequents is impossible.
Indeed, this impossibility has been shown  for the system BV~\cite{Tiu} and apparently applies as well to Pomset logic
(see~\cite{Pogodalla_apomset}), which is believed to be equivalent to BV (inclusion of BV in Pomset logic
is proven in~\cite{StrassburgerThesis}). And Pomset logic is closely related to ScMLL\@.
% particular, for Retor\'e's Pomset logic, which turns out closely related to ScMLL, there are conclusive arguments that an adequate formulation in the language of ordinary sequents is impossible. In~\cite{Tiu} A. Tiu showed that the deep inference system BV, which is believed to be equivalent to  (and definitely is contained in) Pomset logic cannot be captured by a sequent calculus.
%

   On the other hand, for noncommutative logics  it is quite customary to use sequents with an extra structure. For
   example, cyclic linear logic~\cite{Yetter} uses cyclically ordered sequents, and Abrusci-Ruet's noncommutative
    logic~\cite{AbrusciRuet} uses sequents with a complicated additional structure of \emph{order variety}. Retor\'e in his attempts to formulate a sequent calculus for Pomset logic considered partially ordered sequents~\cite{Retore}. Following this tradition we consider sequents decorated with an extra structure of binary relation.

\subsection{Decorated sequents}
We will consider \emph{decorated sequents}, which are sequents equipped with a certain extra structure of \emph{reachability relation}.

In order to make definitions more concise, we introduce the following terminology.
Given a sequent $\Gamma$, a \emph{vector} in $\Gamma$ is any finite nonempty sequence of distinct elements of $\Gamma$. Vectors will systematically be denoted with boldface letters. For a vector ${\bf X}$ we denote the size of ${\bf X}$ as $|{\bf X}|$.
The empty sequence is denoted as
${\bf \Lambda}$.

%\begin{equation}\label{joint_permutation}
%\begin{split}
% {\bf \Lambda} & r {\bf \Lambda}, \\
%   (X_1,\ldots,X_n)r(Y_1\ldots, Y_n)&\mbox{ implies } (X_{\sigma(1)},\ldots,X_{\sigma(n)})r(Y_{\sigma(1)}\ldots, Y_{\sigma(n)}) \\
%  \mbox{ for any permutation }\sigma\in S_n.&
% \end{split}
% \end{equation}

\begin{defi}\label{decorated seq}
A {\bf decorated sequent} is a pair $(\Gamma,\rightrightarrows)$, where $\Gamma$ is a sequent, and $\rightrightarrows$
is a binary relation of {\bf reachability} between vectors in $\Gamma$ satisfying the following properties:
\begin{enumerate}[(i)]
  \item${\bf \Lambda}\rightrightarrows{\bf \Lambda}$;
  \item if ${\bf X}\rightrightarrows{\bf Y}$ then $|{\bf X}|=|{\bf Y}|$;
  \item if $(X_1,\ldots,X_n)\rightrightarrows(Y_1\ldots, Y_n)$ then $(X_{\sigma(1)},\ldots,X_{\sigma(n)})\rightrightarrows(Y_{\sigma(1)}\ldots, Y_{\sigma(n)})$ for any permutation $\sigma\in S_n$;
   \item %\[\mbox{ if }{\bf X}\rightrightarrows{\bf Y}\mbox{ then }|{\bf X}|=|{\bf Y}|;\]
   { if } ${\bf X}\rightrightarrows{\bf Y}$
    { then }
    ${\bf X}$ and ${\bf Y}$ have no common element.
  \end{enumerate}
\end{defi}

\noindent
When ${\bf X}\rightrightarrows{\bf Y}$ we say that \emph{${\bf Y}$ is reachable from ${\bf X}$}.
 If ${\bf Y}$ is not reachable from ${\bf X}$ we write
\[{\bf X}\not\rightrightarrows{\bf Y}.\]

We define also reachability between formulas by considering formulas as single element vectors.
Any decorated sequent whose underlying sequent is $\Gamma$ is called a \emph{decoration} of $\Gamma$. We systematically  abuse notation by denoting both the decoration $(\Gamma,\rightrightarrows)$ and its underlying sequent  as $\Gamma$.

%The intended meaning of the reachability relation is that $XrY$ iff in the corresponding proof-net with  conclusions $\Gamma$, for some switching, in the second round there is a (non-selfintersecting) essentially directed path from the conclusion $X$ to the conclusion $Y$.

%The usage of the reachability relation is that, if $X$ is not reachable from $Y$ (i.e. not $YrX$) in the decorated sequent $\Gamma\cup\{X,Y\}$, then, in the corresponding proof-net we can safely introduce between them a $\vwpop$-link and derive $\Gamma\cup\{X\vwpop Y\}$. Accordingly, in the sequent calculus, if $X$ is not reachable from $Y$ then we can derive $\vdash\Gamma,X\vwpop Y$ from $\vdash\Gamma, X,Y$.

Now let $\rho$ be a proof-net and $\Gamma$ be a decorated sequent.

\begin{defi}\label{proof-net of dec. seq}
We say that $\rho$ is a {\bf proof-net of} $\Gamma$ when the following holds:
\begin{enumerate}[(i)]
  \item the multiset of conclusions of $\rho$ is the underlying sequent of $\Gamma$,
  \item if \[{\bf X}=(X_1,\ldots,X_n),\quad{\bf Y}=(Y_1,\ldots,Y_n)\]
   are vectors in $\Gamma$ with no common element, then it holds that ${\bf X}\rightrightarrows{\bf Y}$ iff
  for some switching of $\rho$ in the second round there exist pairwise nonintersecting essentially directed paths from $X_i$ to $Y_i$, $i=1,\ldots, n$.
\end{enumerate}
\end{defi}

\noindent
A decorated sequent $\Gamma$ \emph{is derivable} in ScMLL if there is a proof-net of  $\Gamma$.

It is easy to see that an ordinary sequent is derivable in ScMLL iff some its decoration is derivable in ScMLL\@.
\begin{rem}\label{some decoration must}
Let $\Gamma$ be an ordinary sequent and $\rho$ be a proof-net of $\Gamma$. Then there exist unique reachability relation $\rightrightarrows$ on vectors in $\Gamma$ such that $\rho$ is a proof-net of the decorated sequent $(\Gamma,\rightrightarrows)$.
\end{rem}

\begin{proof} The relation $\rightrightarrows$ is defined by condition (ii) of Definition~\ref{proof-net of dec. seq}.
\end{proof}

If $\rho$ is a proof-net of an ordinary sequent $\Gamma$ and the reachability relation $\rightrightarrows$ is defined  on $\Gamma$ by condition (ii) of Definition~\ref{proof-net of dec. seq} as in the above proof, then we say that $(\Gamma,\rightrightarrows)$ is the \emph{ decoration of $\Gamma$ in $\rho$}.

\begin{rem}
It can be observed that  reachability relations that actually occur in derivable decorated
sequents are far from being arbitrary. For example, if $\Gamma$ is an  MLL sequent (i.e. $\Gamma$ does not
use $\votimes$ or $\vwpop$ connectives), and $\rho$ is a proof-net of $\Gamma$, then
the decoration of $\Gamma$ in $\rho$ has necessarily \emph{total} reachability relation. That is,
any two vectors in $\Gamma$ of equal size and with no common element are reachable from each other.
\end{rem}

It might be interesting to understand which reachability relations actually do occur. Unfortunately, we cannot say
anything definite on this subject.
\smallskip

The main usage of reachability is the following.

\begin{rem}\label{usage of reachability}
Assume that $\Gamma$ is a decorated sequent and $\rho$ is a proof-net of  $\Gamma$.

Let
%\[{\bf X}=(X_1,\ldots,X_n),\quad{\bf Y}=(Y_1,\ldots,Y_n)\],
%are vectors
$X,Y$ be formulas
of $\Gamma$. Then $Y$  is {\bf not} reachable from $X$ iff the proof-structure $\rho'$ obtained from $\rho$ by attaching  a  $\vwpop$-link between $X$ and $Y$ is a proof-net.
\end{rem}

\begin{proof}
Let $\sigma$ be a switching of $\rho'$.
It is immediate that the first round is connected and acyclic, otherwise the original $\rho$ is not a proof-net.

Assume that  there is an essentially directed cycle $\phi$ in the second round of $\sigma$. Then $\phi$  passes through the conclusion link $X\vwpop Y$, otherwise $\rho$ is not a proof-net. By removing this link from $\phi$ we obtain an essentially directed path from $Y$ to $X$, hence $X\rightrightarrows Y$. The other direction is similar. \end{proof}

%
% The meaning of such an extension is: $(X_1,\ldots,X_n)r(Y_1,\ldots,Y_n)$, where all $X_i,Y_j$, $i,j=1,\ldots,n$, are pairwise distinct elements (i.e. occurrences of formulas) in $\Gamma$, iff in the corresponding proof-net with conclusions $\Gamma$, for some switching, in the second round there exist pairwise nonintersecting essentially directed paths from $X_i$ to $Y_i$, $i=1,\ldots, n$.

\subsubsection{First level of  sequent rules}
Now we want to write a sequent calculus  for decorated sequents.
Every rule will have two levels: the first level is how the underlying sequents are changed, the second, how the reachability relations are changed.

We first list the level of sequents.
Here, the rules are basically, those of ordinary MLL, sometimes with arrows added
 above connectives.
\begin{mathpar}
\inferrule{~}{\vdash A,A^\bot} \; (\mathrm{Axiom})
\and
\inferrule{\vdash\Gamma,A \\ \vdash A^\bot,\Delta}{\vdash \Gamma,\Delta} \; (\mathrm{Cut})
\and
\inferrule{\vdash\Gamma,A \\ \vdash B,\Delta}{\vdash \Gamma,A\otimes B,\Delta} \; (\otimes)
\\
\inferrule{\vdash\Gamma,A,B}{\vdash \Gamma,A\wpop B} \; (\wpop)
\and
\inferrule{\vdash\Gamma,A \and \vdash B,\Delta}{\vdash \Gamma,A \votimes B,\Delta} \; (\votimes)
\and
\inferrule{\vdash\Gamma,A,B\mbox{ }(\mathrm{where}\mbox{ }B\not\rightrightarrows A)}{\vdash \Gamma,A\vwpop B} \; (\vwpop)
\end{mathpar}

Let us make it clear how we read rules on the first level.
For each rule in the above list it is understood that the premises are decorated sequents, whereas the conclusion is an ordinary sequent.

Now we will translate rules in the above list to operations on proof-nets.
More specifically, to each rule
\[
\inferrule{\Theta_1 \and \cdots \and \Theta_n}{\Theta} \; (R)
\]
we assign an operation, which, given proof-nets $\rho_1,\ldots,\rho_n$ of the decorated sequents $\Theta_1,\ldots,\Theta_n$, transforms them into a proof-net $\rho$ of the (ordinary) sequent $\Theta$.

Moreover, \emph{the decoration of $\Theta$ in $\rho$ depends only on $\Theta_1,\ldots,\Theta_n$ and does not depend on $\rho_1,\ldots,\rho_n$}.
Then rules for computing this decoration of $\Theta$ from knowledge of $\Theta_1,\ldots,\Theta_n$ form the second level of $(R)$.
Translation goes as follows.

Axioms translate to proof-links.

If $\rho_1$, $\rho_2$ are proof-nets with conclusions $\Gamma,A$ and $A^\bot,\Delta$ respectively, then putting $\rho_1$, $\rho_2$ together and connecting their respective conclusions $A,A^\bot$ with a Cut-link produces a proof-structure, which is easily seen to be a proof-net. This is the translation of the Cut rule.

Similarly, if $\rho_1$, $\rho_2$ are proof-nets with conclusions $\Gamma,A$ and $B,\Delta$ respectively, then putting $\rho_1$, $\rho_2$ together and connecting their respective conclusions $A,B$ with a $\otimes$-link gives proof-net. This is the translation of the $(\otimes)$ rule. The $(\votimes)$ rule is treated in exactly the same way, with $\otimes$-link changed to $\votimes$-link.

If $\rho$ is a proof-net with conclusions $\Gamma,A,B$, then attaching to conclusions $A,B$ a $\wpop$-link gives a proof-net again. This is the translation of the $(\wpop)$ rule.

Finally, if  $\rho$ is a proof-net of a decorated sequent $\Gamma,A,B$, and $A$ is not reachable from $B$, then by Remark~\ref{usage of reachability}, attaching to conclusions $A,B$ a $\vwpop$-link gives a proof-net as well again. This is the translation of the $(\vwpop)$ rule.

Let us find the second level of rules.

\subsubsection{Second level of the ($\vwpop$) rule}
Let  $\Theta$ be a decorated sequent, whose underlying sequent is $\Gamma,A,B$.
Assume that $\rho$ is a proof-net of $\Theta$, and let $\rho'$ be obtained from $\rho$ by attaching a $\vwpop$-link between $A$ and $B$.
Let \[{\bf X}=(X_1,\ldots,X_n),\quad{\bf Y}=(Y_1,\ldots,Y_n)\] be two vectors in the sequent $\Gamma,A\vwpop B$, which have no element in common.

We need to figure out under which conditions there is a switching of $\rho'$ such that in the second round  there exist  pairwise nonintersecting essentially directed paths $\phi_i$ from $X_i$ to $Y_i$, $i=1,\ldots, n$. We consider possibilities case-by-case.

\begin{itemize}
\item Assume that $A\vwpop B\not\in{\bf X}$, $A\vwpop B\not\in{\bf Y}$, and none of $\phi_1,\ldots,\phi_n$ meets $A$ or $B$.
    Then all $\phi_1,\ldots,\phi_n$ lie entirely in $\rho$, and since $\rho$ is a proof-net of $\Theta$ it follows that ${\bf X}\rightrightarrows{\bf Y}$ in $\Theta$.
\item
Assume that  $A\vwpop B\in{\bf Y}$, and some of  $\phi_1,\ldots,\phi_n$ meets $A$.
So we have $A\vwpop B=Y_j$ for some $j$.
Thus in the second round of $\rho$ there is an essentially directed path from $X_j$ to $A\vwpop B$.
Since  the paths $\phi_1,\ldots,\phi_n$ are pairwise non-intersecting, it follows $\phi_i$ does not meet $A\vwpop B$ unless $i=j$. In particular, none of these paths meets $B$. (If there were such a path, then one of its endpoints would be $A\vwpop B$.)

Then, restricting to $\rho$, we get in $\rho$ an essentially directed path from $X_j$ to $A$.
It follows that
\[(X_1,\ldots,X_{j-1},X_j,X_{j+1},\ldots,X_n)\rightrightarrows(Y_1\ldots,Y_{j-1},A,Y_{j+1},\ldots Y_n)\] in $\Theta$.
\item
 Assume that  $A\vwpop B\in{\bf Y}$, and some of $\phi_1,\ldots,\phi_n$ meets $B$.
Then, just as above, we get
\[(X_1,\ldots,X_{j-1},X_j,X_{j+1},\ldots,X_n)\rightrightarrows(Y_1\ldots,Y_{j-1},B,Y_{j+1},\ldots Y_n)\] in $\Theta$.
\item Assume that  $A\vwpop B\in{\bf X}$, and some of $\phi_1,\ldots,\phi_n$ meets $A$.
Then, just as above, we get
\[(X_1,\ldots,X_{j-1},A,X_{j+1},\ldots,X_n)\rightrightarrows(Y_1\ldots,Y_{j-1},Y,Y_{j+1},\ldots Y_n)\] in $\Theta$.

\item Assume that  $A\vwpop B\in{\bf X}$, and some of $\phi_1,\ldots,\phi_n$ meets $B$.
Then, just as above, we get
\[(X_1,\ldots,X_{j-1}B,X_{j+1},\ldots,X_n)\rightrightarrows(Y_1\ldots,Y_{j-1},Y,Y_{j+1},\ldots Y_n)\] in $\Theta$.

  \item
  Finally, assume that
   $A\vwpop B\not\in{\bf X}$, $A\vwpop B\not\in{\bf Y}$, and some of $\phi_1,\ldots,\phi_n$ meets $A$ or $B$.
If the path $\phi_j$ from $X_j$ to $Y_j$ meets $A$ or $B$, then it meets $A\vwpop B$. Indeed, the endpoints of $\phi_j$ are conclusions of $\rho'$, so neither $A$ nor $B$ can be a conclusion of $\phi_j$.
Furthermore,  the link $A\vwpop B$ is not an endpoint of $\phi_j$, so it is a critical link of $\phi_j$. Then, since $\phi_j$ is essentially directed, it follows that it has the form $X_j\ldots,A,A\vwpop B,B,\ldots,Y_j$.
Then, restricting to $\rho$, we get in $\rho$ essentially directed paths from $X_j$ to $A$ and from $B$ to $Y_j$.
It follows that
\[(X_1,\ldots,X_j,B,X_{j+1},\ldots,X_n)\rightrightarrows(Y_1\ldots,Y_{j-1},A,Y_j\ldots Y_n)\] in $\Theta$.
\item There are no other possibilities.
 \end{itemize}

\noindent
  The decoration of  $\Gamma,A\vwpop B$ in $\rho'$  is read from the above exhausting list.

  Using property (iii) in the definition~\ref{decorated seq}, the rules for computing the decoration can be written in a more compact form.

{\bf Rule $(\vwpop)$, second level:}

  Reachability on $\Gamma,A\vwpop B$ is  the smallest relation satisfying properties (i)--(iv) of Definition~\ref{decorated seq} and the following conditions: % chktex 36
\begin{itemize}
  \item If ${\bf X}\rightrightarrows{\bf Y}$ in  $\Theta$, then ${\bf X}\rightrightarrows{\bf Y}$ in  $\Theta'$.
  \item If $({\bf X'},X)\rightrightarrows({\bf Y},A)$ or $({\bf X'},X)\rightrightarrows({\bf Y},B)$ in  $\Theta$, then  $({\bf X'},X)\rightrightarrows({\bf Y},A\vwpop B)$ in  $\Theta'$. Here $A,B\not\in({\bf X'},{\bf Y},X)$.
  \item  If $({\bf X},A)\rightrightarrows({\bf Y'},Y)$ or $({\bf X},B)\rightrightarrows({\bf Y'},Y)$ in  $\Theta$, then $({\bf X},A\vwpop B)\rightrightarrows({\bf Y'},Y)$ in  $\Theta'$. Here $A,B\not\in({\bf X},{\bf Y'},Y)$.

  \item  If $({\bf X'},X,B)\rightrightarrows({\bf Y'},A,Y) $ in  $\Theta$, then $({\bf X'},X)\rightrightarrows({\bf Y'},Y)$ in $\Theta'$.
\end{itemize}

%Note that definition of $\Theta'$ does not depend on $\rho$, but only on reachability relation in $\Theta$.
%
%From the above discussion we obtain the following.
%
%\begin{lem}\label{vecwp rule}
%In notations as above, if $\Theta$  is provable in ScMLL, then $\Theta'$ is provable in ScMLL
% as well. \qed%
% \end{lem}

\subsubsection{Second level of sequent rules}
Second level for other rules is computed by a case-by-case analysis as in the preceding section.
We simply write down the result.
For each sequent  rule, reachability relation on the conclusion is the smallest relation satisfying properties (i)--(iv) and the following conditions. % chktex 36

\begin{description}
\item[The axioms]
 $A\rightrightarrows A^\bot$, $A^\bot \rightrightarrows A$.

\item[The $\vwpop$-rule]
If ${\bf X}\rightrightarrows{\bf Y}$ in the premise, then ${\bf X}\rightrightarrows{\bf Y}$ in the conclusion.

 If $({\bf X'},X)\rightrightarrows({\bf Y},A)$ or $({\bf X'},X)\rightrightarrows({\bf Y},B)$ in the premise, then in the conclusion $({\bf X'},X)\rightrightarrows({\bf Y},A\vwpop B)$. Here $A,B\not\in({\bf X'},{\bf Y},X)$.

 If $({\bf X},A)\rightrightarrows({\bf Y'},Y)$ or $({\bf X},B)\rightrightarrows({\bf Y'},Y)$ in the premise, then $({\bf X},A\vwpop B)\rightrightarrows({\bf Y'},Y)$ in the conclusion. Here $A,B\not\in({\bf X},{\bf Y'},Y)$.

 If $({\bf X'},X,B)\rightrightarrows({\bf Y'},A,Y) $ in the premise then $({\bf X'},X)\rightrightarrows({\bf Y'},Y)$ in the conclusion.

% {\bf Remark} Note that it this rule where we need to extend reachability relations from  formulas to sequences of  formulas. For example, in order to decide whether $XrY$ in the conclusion we need to know if $(X,B)\rightrightarrows(A,Y)$ in the premise.

\item[The $\otimes$-rule]
If we have in the first premise ${\bf X_1}\rightrightarrows{\bf Y_1}$, and in the second premise ${\bf X_2}\rightrightarrows{\bf Y_2}$, then in the conclusion $({\bf X_1},{\bf X_2})\rightrightarrows({\bf Y_1},{\bf Y_2})$. Here we suppose $A\not\in ({\bf X_1}, {\bf Y_1})$, $B\not\in ({\bf X_2}, {\bf Y_2})$.

If we have in the first premise $({\bf X_1},X)\rightrightarrows({\bf Y_1},A)$, and in the second premise $({\bf X_2},B)\rightrightarrows({\bf Y_2},Y)$, then in the conclusion $({\bf X_1},{\bf X_2},X)\rightrightarrows({\bf Y_1},{\bf Y_2},Y)$. Here we suppose $X\not\in ({\bf X_1}, {\bf Y_1})$, $Y\not\in ({\bf X_2}, {\bf Y_2})$.

Similarly, if in the first premise $({\bf X_1},A)\rightrightarrows({\bf Y_1},Y)$, and in the second premise $({\bf X_2},X)\rightrightarrows({\bf Y_2},B)$, then in the conclusion $({\bf X_1},{\bf X_2},X)\rightrightarrows({\bf Y_1},{\bf Y_2},Y)$. Here we suppose $Y\not\in ({\bf X_1}, {\bf Y_1})$, $X\not\in ({\bf X_2}, {\bf Y_2})$

If we have in the first premise $({\bf X_1},X)\rightrightarrows({\bf Y_1},A)$, and in the second premise ${\bf X_2}\rightrightarrows{\bf Y_2}$, then in the conclusion $({\bf X_1},{\bf X_2},X)\rightrightarrows({\bf Y_1},{\bf Y_2},A\otimes B)$.  Here we suppose   $X\not\in ({\bf X_1}, {\bf Y_1})$, $B\not\in ({\bf X_2}, {\bf Y_2})$.

If  in the first premise ${\bf X_1}\rightrightarrows{\bf Y_1}$, and in the second premise $({\bf X_2},X)\rightrightarrows({\bf Y_2},B)$, then in the conclusion $({\bf X_1},{\bf X_2},X)\rightrightarrows({\bf Y_1},{\bf Y_2},A\otimes B)$.  Here we suppose   $A\not\in ({\bf X_1}, {\bf Y_1})$, $X\not\in ({\bf X_2}, {\bf Y_2})$.

Similarly, if in the first premise $({\bf X_1},A)\rightrightarrows({\bf Y_1},Y)$, and in the second premise ${\bf X_2}\rightrightarrows{\bf Y_2}$, then in the conclusion $({\bf X_1},{\bf X_2},A\otimes B)\rightrightarrows({\bf Y_1},{\bf Y_2},Y)$.  Here we suppose   $Y\not\in ({\bf X_1}, {\bf Y_1})$,  $B\not\in ({\bf X_2}, {\bf Y_2})$.

If in the first premise ${\bf X_1}\rightrightarrows{\bf Y_1}$, and in the second premise $({\bf X_2},B)\rightrightarrows({\bf Y_2},Y)$, then in the conclusion $({\bf X_1},{\bf X_2},A\otimes B)\rightrightarrows({\bf Y_1},{\bf Y_2},Y)$.  Here we suppose  $A\not\in ({\bf X_1}, {\bf Y_1})$,  $Y\not\in ({\bf X_2}, {\bf Y_2})$.

\item[The $\wpop$-rule]
 If ${\bf X}\rightrightarrows{\bf Y}$ in the premise, then ${\bf X}\rightrightarrows{\bf Y}$ in the conclusion.

 If $({\bf X'},X)\rightrightarrows({\bf Y},A)$ or $({\bf X'},X)\rightrightarrows({\bf Y},B)$ in the premise, then  in the conclusion $({\bf X'},X)\rightrightarrows({\bf Y},A\wpop B)$. Here $A,B\not\in({\bf X'},{\bf Y},X)$.

 If $({\bf X},A)\rightrightarrows({\bf Y'},Y)$ or $({\bf X},B)\rightrightarrows({\bf Y'},Y)$ in the premise, then $({\bf X},A\wpop B)\rightrightarrows({\bf Y'},Y)$ in the conclusion. Here $A,B\not\in({\bf X},{\bf Y'},Y)$.

\item[The $\votimes$-rule]
See the preceding section.
%If we have in the first premise ${\bf X_1}\rightrightarrows{\bf Y_1}$, and in the second premise ${\bf X_2}\rightrightarrows{\bf Y_2}$, then in the conclusion $({\bf X_1},{\bf X_2})\rightrightarrows({\bf Y_1},{\bf Y_2})$. Here we suppose $A\not\in ({\bf X_1}, {\bf Y_1})$, $B\not\in ({\bf X_2}, {\bf Y_2})$.

If we have in the first premise $({\bf X_1},X)\rightrightarrows({\bf Y_1},A)$, and in the second premise $({\bf X_2},B)\rightrightarrows({\bf Y_2},Y)$, then we have  $({\bf X_1},{\bf X_2},X)\rightrightarrows({\bf Y_1},{\bf Y_2},Y)$ in the conclusion. Here we suppose $Y\not\in ({\bf X_2}, {\bf Y_2})$, $X\not\in ({\bf X_1}, {\bf Y_1})$.

If we have in the first premise $({\bf X_1},X)\rightrightarrows({\bf Y_1},A)$, and in the second premise ${\bf X_2}\rightrightarrows{\bf Y_2}$, then in the conclusion $({\bf X_1},{\bf X_2},X)\rightrightarrows({\bf Y_1},{\bf Y_2},A\votimes B)$.  Here we suppose   $X\not\in ({\bf X_1}, {\bf Y_1})$.

If in the first premise ${\bf X_1}\rightrightarrows{\bf Y_1}$, and in the second premise $({\bf X_2},X)\rightrightarrows({\bf Y_2},B)$, then in the conclusion $({\bf X_1},{\bf X_2},X)\rightrightarrows({\bf Y_1},{\bf Y_2},A\votimes B)$.  Here we suppose  $A\not\in ({\bf X_1}, {\bf Y_1})$,  $X\not\in( {\bf X_2}, {\bf Y_2})$.

If we have in the first premise $({\bf X_1},A)\rightrightarrows({\bf Y_1},Y)$, and in the second premise ${\bf X_2}\rightrightarrows{\bf Y_2}$, then in the conclusion $({\bf X_1},{\bf X_2},A\votimes B)\rightrightarrows({\bf Y_1},{\bf Y_2},Y)$.  Here we suppose   $Y\not\in( {\bf X_1}, {\bf Y_1})$,  $B\not\in ({\bf X_2}, {\bf Y_2})$.

If  in the first premise ${\bf X_1}\rightrightarrows{\bf Y_1}$, and in the second premise $({\bf X_2},B)\rightrightarrows({\bf Y_2},Y)$, then in the conclusion $({\bf X_1},{\bf X_2},A\votimes B)\rightrightarrows({\bf Y_1},{\bf Y_2},Y)$.   We suppose here    $A\not\in ({\bf X_2}, {\bf Y_2})$, $Y\not\in ({\bf X_2}, {\bf Y_2})$.

\item[The Cut rule]
If we have in the first premise $({\bf X_1},X)\rightrightarrows({\bf Y_1},A)$, and in the second premise $({\bf X_2},A^\bot)\rightrightarrows({\bf Y_2},Y)$, then in the conclusion $({\bf X_1},{\bf X_2},X)\rightrightarrows({\bf Y_1},{\bf Y_2},Y)$. Here we suppose $Y\not\in {\bf X_2}, {\bf Y_2}$, $X\not\in ({\bf X_1}, {\bf Y_1})$.

Similarly, if in the first premise $({\bf X_1},A)\rightrightarrows({\bf Y_1},Y)$, and in the second premise $({\bf X_2},X)\rightrightarrows({\bf Y_2},A^\bot)$, then in the conclusion $({\bf X_1},{\bf X_2},X)\rightrightarrows({\bf Y_1},{\bf Y_2},Y)$. Here we suppose $Y\not\in {\bf X_1}, {\bf Y_1}$, $X\not\in ({\bf X_2}, {\bf Y_2})$.
\end{description}
\smallskip

\noindent
Now, by the very construction, any decorated sequent calculus derivation of a decorated sequent $\Gamma$ translates into a proof-net of $\Gamma$. This implies \emph{soundness} of the (decorated) sequent calculus.

\begin{lem}\label{soundness}
If a decorated sequent $\Theta$ is derivable in the decorates sequent calculus above, then it is derivable in ScMLL\@.
\end{lem}
%{\bf Proof} Given a derivation $\sigma$ in the decorated sequent calculus, we translate it step by step into a proof-net $\rho(\sigma)$, using translations of rules discussed above.
%
%Then we prove by induction on $\sigma$ that if $\vdash\Theta$ is the conclusion of $\sigma$, then $\rho(\sigma)$ is a proof-net of $\Theta$. This is done

\subsection{Completeness}
Now we are going to prove  completeness of the decorated sequent calculus.
Similarly to the case of proof-nets there is the \emph{commutative projection} map from decorated sequent calculus proofs to ordinary MLL sequent calculus proofs, obtained by erasing all arrows above connectives, and forgetting decorations.
In more details, and restricting to cut-free proofs:
\begin{itemize}
  \item an ScMLL formula $F$ is mapped to an MLL formula $F'$ by replacing each $\votimes$, respectively, $\vwpop$ connective to $\otimes$, respectively, $\wpop$ connective;
  \item an ScMLL sequent $\Theta=F_1,\ldots,F_n$ is mapped to the MLL sequent $F_1',\ldots,F_n'$;
  \item a decorated sequent calculus proof $\sigma$ of $\Gamma$ is mapped to an MLL sequent calculus proof $\sigma'$ of $\Gamma'$ by induction. Axiom is mapped to itself. A decorated sequent calculus proof obtained from proofs $\sigma_1$, $\sigma_2$ by the $(\otimes)$ or $(\votimes)$ rule is mapped to the MLL proof obtained from $\sigma_1$, $\sigma_2$ by the $(\otimes)$ rule.  A decorated sequent calculus proof obtained from a proof $\sigma_0$ by the $(\wpop)$ or $(\vwpop)$ rule is mapped to the MLL proof obtained from $\sigma_0$ by the $(\wpop)$ rule.
\end{itemize}

\noindent
Also, recall from the preceding section that any decorated sequent calculus derivation translates to an ScMLL proof-net. Similarly, an MLL sequent derivation translates to an MLL proof-net. In both cases we denote the proof-net obtained from a derivation $\sigma$ as $T(\sigma)$.

%\begin{lem} There is a map from ScMLL sequent proofs to proof-nets which sends a proof $\pi$ of the decorated sequent $\Gamma$ to a proof-net $\rho(\pi)$ with conclusions $\Gamma$, such that for all sequences ${\bf X}=( X_1,\ldots,X_n),{\bf Y}=(Y_1,\ldots, Y_n)\subset\Gamma$, where all $X_i,Y_j$, $i,j=1,\ldots,n$, are pairwise distinct, it holds that ${\bf X}r_\Gamma {\bf Y}$  iff for some  switching of $\rho(\pi)$, in the second round there are pairwise nonintersecting non-selfintersecting essentially directed paths from the conclusion
% $X_i$ to the conclusion $Y_i$, $i=1,\ldots,n$.
%\end{lem}
%{\bf Proof} The map is defined by induction, starting from the axioms which are mapped to axiom links.
% \smallskip

\begin{lem}\label{sequentialization} There is a {\bf sequentialization} map from cut-free proof-nets to cut-free ScMLL sequent proofs
 which sends a proof-net $\rho$ of a decorated sequent  $\Theta$ to a decorated  sequent calculus derivation $\pi(\rho)$
 of $\Theta$.
   \end{lem}

 \begin{proof}
 There exists a sequentialization map sending a cut-free MLL proof-net of a sequent $\Gamma$ to a cut-free MLL derivation of $\Gamma$. Let us denote this map as $\pi$. Then for any MLL proof-net $\rho$ we have $T\circ\pi(\rho)=\rho$.

 Now let $\rho$ be a cut-free  ScMLL proof-net of a decorated sequent $\Gamma$, and let  $\rho'$ be its commutative projection.
  Then we have an MLL sequent calculus derivation $s=\pi(\rho')$ of  $\Theta'$.
  By induction on $s$ we show that there exists a decorated sequent calculus derivation $\sigma$ of $\Theta$ such that the commutative projection $\sigma'=s$ and the translation $T(\sigma)=\rho$.

  If $s$ is an axiom, then $\rho=\rho'$ is an axiom link, and we take $\sigma =s$.

 Let  $s$ be obtained from an MLL derivation $s_0$ by the  $(\wpop)$ rule as follows
 \[\inferrule{\vdash\Gamma_0,A_0,B_0}{\vdash\Gamma_0,A_0\wpop B_0} \; (\wpop).\]
 Then the MLL proof-net $T(s)=\rho'$ is obtained from $T(s_0)$ by attaching an $\wpop$-link. It follows that
 $\Theta'=\Gamma_0,A\wpop B$ and
 \begin{equation}\label{1st case}
 \Theta=\Gamma,A\wpop B
  \end{equation}
  or
  \begin{equation}\label{2nd case}
  \Theta=\Gamma,A\vwpop B
  \end{equation} for some $\Gamma,A,B$ such that $\Gamma_0=\Gamma'$, $A_0=A'$, $B_0=B'$.

 By induction hypothesis there exists a decorated sequent calculus derivation $\sigma_0$ of $\Gamma,A,B$.
 If  (\ref{1st case}) holds, we obtain $\sigma$ from $\sigma_0$ by the $(\wpop)$ rule.
 If (\ref{2nd case}) holds, then formulas $A$, $B$ are connected with a $\vwpop$-link in $\rho$. Then by Remark~\ref{usage of reachability} $A$ is not reachable from $B$, and we can apply the $(\vwpop)$ rule to $\sigma_0$, which gives the desired $\sigma$.
Remaining steps are similar.

 We define $\pi(\rho)=\sigma$. \end{proof}

\section{Variations of the logic}
We now proceed to ``degenerate'' variations of ScMLL\@. Our final goal is to obtain a version with self-dual $\vwpop$, which turns out equivalent to Pomset logic.

 \subsection{Adding Mix}
 Recall that the system MLL+Mix is  obtained from MLL by adding the \emph{Mix rule}
 \begin{equation}\label{Mix}
  \inferrule{\vdash\Gamma\mbox{ }\vdash\Delta}{\vdash\Gamma,\Delta} \; (\mbox{Mix}).
\end{equation}
See~\cite{FR} for a syntax of proof-nets.

We want  to add this rule to ScMLL to get the system ScMLL+Mix.
  The \emph{proof-structures} for ScMLL+Mix are the same as for ScMLL\@. The definition of  proof-nets is relaxed in the first-round part, right as in the familiar case of Danos-Regnier criterion for MLL+Mix~\cite{FR}.
   \begin{defi}\label{SCMLL+Mix proof-net} An ScMLL+Mix proof-net is a proof-structure $\rho$, such that for any switching $\sigma$ the first round is  acyclic, and the second round has no essentially directed cycles.
   \end{defi}

The decorated \emph{sequent calculus} for ScMLL+Mix is also the same as for ScMLL, supplied with a decorated version of  Mix rule.

\begin{defi}
In the decorated sequent calculus the Mix rule is defined by (\ref{Mix}) supplied with the following second level part.

{\rm Reachability relation on the conclusion of (\ref{Mix}) is the smallest relation satisfying properties (i)--(iv) of Definition~\ref{decorated seq} and the condition % chktex 36
\begin{itemize}
  \item  if ${\bf X_1}\rightrightarrows{\bf Y_1}$ in the first premise, and   ${\bf X_2}\rightrightarrows{\bf Y_2}$ in the second  premise, then   $({\bf X_1},{\bf X_2})\rightrightarrows({\bf Y_1},{\bf Y_2})$ in the conclusion.
\end{itemize}
}
\end{defi}

\noindent
All above constructions and results for ScMLL go as well for ScMLL with minor variations, which are straightforward.

We summarise in the  following
 \begin{thm}
 There is an algorithm transforming any ScMLL proof-net with cut-links to a cut-free ScMLL proof-net of the  same sequent.

   A sequent is derivable in ScMLL+Mix iff it is derivable by means of a cut-free proof-net.

   ScMLL+Mix is a conservative extension of MLL+Mix.
   A sequent $\Gamma$ is derivable in ScMLL+Mix iff some its decoration is derivable in ScMLL+Mix sequent calculus iff some its decoration is derivable in the cut-free ScMLL+Mix sequent calculus. \qed%
 \end{thm}
\smallskip

Observe that using the Mix rule we can derive not only the ``commutative'' Mix principle \[A\otimes B\vdash A\wpop B,\] but its ``noncommutative'' version as well
\begin{equation}\label{noncommutative mix}
 A\votimes B\vdash A\vwpop B.
 \end{equation}
A proof-net for (\ref{noncommutative mix}) is the following one.
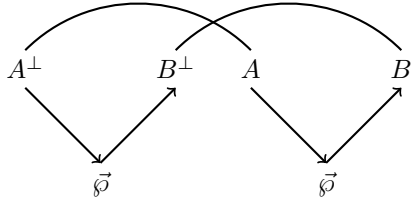
\[
\begin{tikzpicture}{scale =2}
\node[above] at (0,0) {$A^\bot$};
\node[above] at (2,0) {$B^\bot$};
\draw[thick,->] (0,0) --(1,-1);
\draw [thick,<-](2,0) --(1,-1);
\node[below] at (1,-1) {$\vwpop$};
\node[above] at (3,0) {$A$};
\node[above] at (5,0) {$B$};
\draw[thick,->] (3,0) --(4,-1);
\draw [thick,<-](5,0) --(4,-1);
\node[below] at (4,-1) {$\vwpop$};
\draw [thick](0,0.5) to [out=45,in=135] (3,0.5);
%\node[above] at (2.5,1){Id};
\draw [thick](2,0.5) to [out=45,in=135] (5,0.5);
\end{tikzpicture}
\]

Observe that this is not a proof-net for ScMLL, because the first round of any of its switchings is not connected.

\smallskip
In the spirit of Remark~\ref{what_is_provable} let us summarize.
\begin{rem} The logic ScMLL+Mix contains ScMLL\@.
   ScMLL+Mix derives (\ref{n.c.seq_with_mix}).

   ScMLL+Mix still does not derive $A\votimes B\vdash B\votimes A$. \qed%
   \end{rem}

\subsection{Self-dual \texorpdfstring{$\vwpop$}{phi}. Retor\'e logic}
Having Mix, we can add one more ``level of degeneracy'' and obtain a consistent system with $\votimes$ and $\vwpop$ declared isomorphic. This system turns  out equivalent to Pomset logic of Retor\'e, and it seems fair to call it \emph{Retor\'e logic} (RL).

Now, having two distinct connectives $\votimes$ and $\vwpop$ which are exactly equivalent leads to very
uneconomical definitions, where lot of things are just repeated twice.
Therefore we define the \emph{language of RL} as that of MLL supplied with \emph{one} additional connective $\vwpop$.

Furthermore we \emph{redefine} {linear negation} for RL\@.
\emph{Linear negation} in RL is defined as in MLL, supplied with the rule
\[{(A \vwpop B)}^\bot=A^\bot\vwpop B^\bot.\]

\begin{rem}
The reason we keep notation $\vwpop$ for the unique noncommutative connective of RL is that we want to use commutative projections, just as in the case of ScMLL\@. And projecting the noncommutative connective  to $\wpop$ produces correct proof-nets for MLL+Mix.
\end{rem}

\emph{Proof-structures} for RL are defined formally as proof-structures for ScMLL without $\votimes$-links. We should keep in mind, however, that RL proof-structures have cut-links \emph{different} from ScMLL proof-structures, because the negation is different.
Indeed, for a formula $A=F\vwpop G$, the cut-link
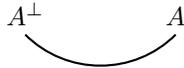
\[
\begin{tikzpicture}{scale =2}
\node[above] at (0,0) {$A^\bot$};
\node[above] at (2,0) {$A$};
\draw [thick](0,0) to [out=-45,in=-135] (2,0);
%\draw[thick,->](0,0)--(1,-1);
%\draw[thick,->](1,-1)--(2,0);
%\node[below] at (1,-1){Cut};
\end{tikzpicture}
\]
looks as the following,
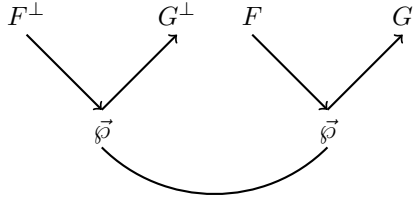
\[
\begin{tikzpicture}{scale =2}\label{cut in RL}
\node[above] at (0,0) {$F^\bot$};
\node[above] at (2,0) {$G^\bot$};
\draw [thick](1,-1.6) to [out=-45,in=-135] (4,-1.6);
\draw[thick,->](0,0)--(1,-1);
\draw[thick,->](1,-1)--(2,0);
\node[below] at (1,-1){$\vwpop$};
\node[above] at (3,0) {$F$};
\node[above] at (5,0) {$G$};
\draw[thick,->](3,0)--(4,-1);
\draw[thick,->](4,-1)--(5,0);
\node[below] at (4,-1){$\vwpop$};
%\draw[thick](1,-1.5)--(2.5,-2.5)--(4,-1.5);
%\node[below] at (2.5,-2.5){Cut};
\end{tikzpicture}
\]
and such a link does not exist in ScMLL\@.

With this in mind, definition of an \emph{RL proof-net} is the same as Definition~\ref{SCMLL+Mix proof-net} of ScMLL+Mix proof-net, with words ``ScMLL+Mix'' replaced with ``RL''.

The \emph{sequent calculus} is defined formally by the rules of ScMLL+Mix in the language without $\votimes$ (hence there is no $\votimes$ rule). But, again, we should keep in mind that individual instances of the Cut-rule are \emph{different}  from those in ScMLL+Mix, because the negation is different.

The results and constructions for ScMLL and ScMLL+Mix go for RL as well with minor variations. In particular, the correspondence between cut-free proof-nets and cut-free sequent proofs is obtained for free, because the cut-free part of RL (both on the proof-nets side and on the decorated sequents side) is \emph{literally} the same as the cut-free part of  ScMLL+Mix in the language without $\votimes$.
 The only point that may deserve some attention is cut-elimination for proof-nets.

\begin{lem} There is an algorithm transforming any RL proof-net with cut-links to a cut-free RL proof-net with the same
   conclusions.
   \end{lem}

   \begin{proof} Again, the algorithm is that of ordinary MLL as far as unordered links are concerned.

The new case is when the cut-link is between formulas $ A\vwpop B$,
   $A^\bot\vwpop B^\bot$.
   This is replaced by two cut-links, between $A$ and $A^\bot$ and between $B$ and $B^\bot$.

   We need to check that this step transforms proof-nets to proof-nets. But now we should consider the first-round part
   as well: we cannot refer to commutative projections, because there is no such a reduction step in the commutative logic.

    Thus,  let $\rho$ be an RL proof-net with a cut-link between $ A\vwpop B$ and
   $A^\bot\vwpop B^\bot$, and let the proof-structure $\rho'$
   be obtained from $\rho$ by a one-step reduction.

     Assume that for some switching $\sigma$ of $\rho'$ there is a cycle $\phi$ in the
   first round.
   Observe then that $\phi$ has no critical links (indeed, there are no $\votimes $-links, and for any
   $\vwpop$-link, one edge is erased by $\sigma$). Hence $\phi$ is essentially directed, even stronger: it is
    essentially directed for both possible choices of its direction. Then, removing from $\phi$ the cut-links between
    $A,A^\bot$ and $B,B^\bot$ (the path $\phi$ must pass through at least one of these links) we obtain at least one
     essentially directed path $\phi_0$ in $\rho\cap\rho'$, connecting two vertices in the set $A,A^\bot,B,B^\bot$.
     But this gives rise to an essentially directed cycle in the second round for any switching of $\rho$ that agrees
     with $\sigma$ on $\rho\cap\rho'$. Which is impossible.

   The second-round part goes exactly as in Lemma~\ref{ScMLL cut-reduction} for ScMLL\@.
   \end{proof}

Again we summarize with a theorem.
\begin{thm}
A sequent is derivable in RL iff it is derivable by means of a cut-free proof-net.

RL is a conservative extension of MLL+Mix.

   A sequent $\Gamma$ is derivable in RL iff some its decoration is derivable in RL sequent calculus iff some its decoration is derivable in the cut-free RL sequent calculus.
    \end{thm}

 In the next section we show that RL is equivalent to Pomset logic. As a consequence, RL decorated sequent calculus gives a sequent calculus formulation of Pomset logic, which is one of the key results of the paper.

\section{Relationship with Pomset logic}
\subsection{R\&B proof-nets and Pomset logic}
We define \emph{Pomset logic}~\cite{Retore}  in the language of RL, i.e. MLL plus one self-dual connective $\vwpop$, by means of special proof-nets.
(Retor\'e in~\cite{Retore} uses  notation $<$ for the noncommutative connective)

Again proof-nets are defined in two steps.

An \emph{R}\&\emph{B proof-structure} is a mixed graph whose edges have one of the two types, \emph{regular} or \emph{bold}, some regular edges may be directed, and whose vertices are labeled by formulas, connectives or symbols ``Cut''. Some  vertices, labeled by formulas are \emph{conclusions}.

%{\bf Remark} The ``R\&B'' in the title refers to ``red'' and ``blue'': in Retor\'e's original definition the two types of edges have different colours. For typesetting reasons we change this opposition to ``regular'' and ``bold''.
%\smallskip

Proof-structures are built inductively by the following rules.

A graph with two vertices labeled by dual propositional symbols $p$ and $p^\bot$ connected by one undirected bold edge is a proof-structure with conclusions $p$, $p^\bot$,
  an \emph{axiom link}.
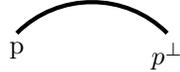
\[
\begin{tikzpicture}
\draw [ultra thick](0,0) to [out=45,in=135] (2,0);
%\node at (1,0.7) {Id};
\node[below] at (0,0) {\strut $p$};
\node[below] at (2,0) {\strut $p^\bot$};
\end{tikzpicture}
\]

A disjoint union of two proof-structures $\rho_1$ and $\rho_2$ is a proof-structure whose conclusions are those of $\rho_1$ or $\rho_2$.

If $\rho$ is a proof-structure whose conclusions contain two (vertices labeled by) formulas $A$ and $B$, then a new proof-structure $\rho'$ can be obtained, by attaching to $A$ and $B$  one of the following graphs,

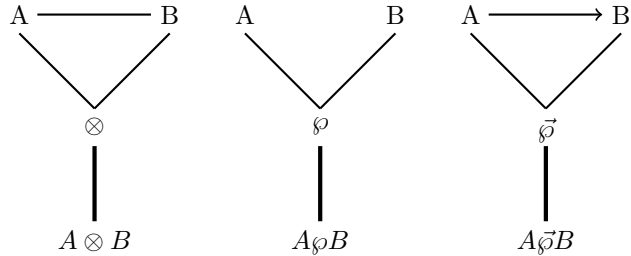
\[
\begin{tikzpicture}{scale =2}
\node[above] at (0,0) {A};
\node[above] at (2,0) {B};
\draw [ thick] (.24,.25) --(1.75,.25);
\draw [ thick] (0,0) --(1,-1);
\draw [ thick] (2,0) --(1,-1);
\node[below] at (1,-1) {$\otimes$};
\draw [ultra thick] (1,-1.5) --(1,-2.5);
\node[below] at (1,-2.5) {$A\otimes B$};

\node[above] at (3,0) {A};
\node[above] at (5,0) {B};
\draw [ thick] (3,0) --(4,-1);
\draw [ thick] (5,0) --(4,-1);
\node[below] at (4,-1) {$\wpop$};
\node[below] at (4,-2.5) {$A\wpop B$};
\draw[ultra thick] (4,-1.5) --(4,-2.5);

\node[above] at (6,0) {A};
\node[above] at (8,0) {B};
\draw [ thick,->] (6.24,.25) --(7.75,.25);
\draw [ thick] (6,0) --(7,-1);
\draw [ thick] (8,0) --(7,-1);
\node[below] at (7,-1) {$\vwpop$};
\draw [ultra thick] (7,-1.6) --(7,-2.5);
\node[below] at (7,-2.5) {$A\vwpop B$};
\end{tikzpicture}
\]
respectively, $\otimes$-link, $\wpop$-link, and $\vwpop$-link.

In each of the above links, the edge connecting the vertex labeled by the  connective to the vertex labeled by the conclusion is bold, and all other edges are regular. In $\otimes$- and $\wpop$-link all edges are undirected, and in the $\vwpop$-link with conclusion $A\vwpop B$, the only directed edge is the regular edge from $A$ to $B$.
The new proof-structure $\rho'$ has the same conclusions as $\rho$ except for $A$ and $B$, and one new conclusion, respectively $A\otimes B$, $A\wpop B$, or $A\vwpop B$, depending on the link attached.

If $\rho$ is a proof-structure whose conclusions contain two (vertices labeled by) dual formulas $A$ and $A^\bot$, then a new proof-structure $\rho'$ can be obtained, by attaching the following graph,  a \emph{Cut-link},
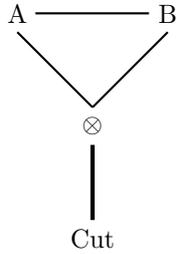
\[
\begin{tikzpicture}{scale =2}
\node[above] at (0,0) {A};
\node[above] at (2,0) {B};
\draw [thick] (.24,.25) --(1.75,.25);
\draw [ thick] (0,0) --(1,-1);
\draw [ thick] (2,0) --(1,-1);
\node[below] at (1,-1) {$\otimes$};
\draw [ultra thick] (1,-1.5) --(1,-2.5);
\node[below] at (1,-2.5) {Cut};
\end{tikzpicture}
\]
which is similar to a $\otimes$-link.

The new proof-structure $\rho'$ has the same conclusions as $\rho$ except for $A$ and $A^\bot$.

An \emph{alternating path} in the proof-structure $\rho$ is a directed path
 whose edges alternate: \ldots-regular-bold-regular-\ldots
An \emph{alternating elementary circuit} (\emph{a.e.\ circuit})  in the proof-structure $\rho$ is an alternating path whose
endpoint vertices coincide, and which traverses all its other vertices exactly once.
%Following terminology of the current paper, an  a.e. circuit is just an alternating directed cycle.

\begin{defi} An R{\rm{\&}}B proof-net is a proof-structure that has no a.e.\ circuit.
\end{defi}

\begin{defi} A sequent $\Gamma$ is derivable in Pomset logic if there exists an R{\rm{\&}}B proof-net whose multiset of conclusions is $\Gamma$. A formula $A$ is derivable in Pomset logic if there exists an R{\rm{\&}}B proof-net whose only conclusion is $A$.
\end{defi}

%(Original Retor\'e definition allows more general proof-nets, with a partial order on conclusions, and sequents with a partial order on formulas. We do not need this generality, because a logic eventually is defined by the set of provable \emph{formulas}.)

The following is proven in~\cite{Retore}.
\begin{thmC}[\cite{Retore}]
A sequent is derivable in Pomset logic iff it is derivable by means of a cut-free R{\rm{\&}}B proof-net.

Pomset logic is a conservative extension of MLL+Mix. \qed%
\end{thmC}
\smallskip

We are going to prove that Pomset logic is equivalent to RL\@. For that purpose, in the next section we introduce alternative proof-nets for RL,  which make this equivalence more transparent.

\subsection{Retor\'e proof-nets for Retor\'e logic}\label{retore proof-nets}
It turns out that proof-nets for RL can be defined in a simpler way, using switchings with only one round.

\begin{defi} A Retor\'e switching of an RL proof-structure is a graph obtained by:
\begin{enumerate}
    \item for each $\wpop$-link, erasing one of the two edges forming the link,
    \item for some of $\vwpop$-links, erasing one of the two edges forming the link and making the remaining edge undirected.
\end{enumerate}
\end{defi}
\begin{defi} A Retor\'e proof-net is an RL proof-structure such that for any Retor\'e switching the resulting graph has no elementary
directed cycles.
\end{defi}

\begin{lem} An RL proof-structure $\rho$ is an RL proof-net iff it is a  Retor\'e proof-net.
\end{lem}

\begin{proof} Assume that for some Retor\'e switching of $\rho$ there is an elementary directed cycle $\phi$. Define an ordinary switching $\sigma$ of $\rho$ by choosing, for each  $\wpop$-link traversed by $\phi$ and for each $\vwpop$-link traversed by $\phi$ and not critical for $\phi$, the edge that is not traversed by $\phi$ and extending this to the whole of $\rho $ in an arbitrary way. If $\phi$
has no critical links, then  $\phi$ is a cycle in the first round of $\sigma$. Otherwise $\phi$ is an essentially directed cycle in the second round. Hence $\rho$ is not a proof-net

Assume that $\rho$ is not a proof-net. Then there exists a switching $\sigma$ of $\rho$ with either a cycle in the first round, or an essentially directed cycle in the second round.

If there is a cycle $\phi$ in the first round, then define a Retor\'e switching by erasing all edges, chosen by $\sigma$. The cycle $\phi$ is directed for this Retor\'e switching (because it traverses only undirected edges).

If there is an essentially directed cycle in the second round,  define a Retor\'e switching by erasing, for each  $\wpop$-link and for each $\vwpop$-link  not critical for $\phi$, the edge chosen by $\sigma$. The cycle $\phi$ is directed in this Retor\'e switching.
Hence $\rho$ is not a Retor\'e proof-net.
\end{proof}

\begin{cor} A sequent $\Gamma$ is derivable  in RL iff there exists a  Retor\'e proof-net with conclusions $\Gamma$ iff there exists a cut-free Retor\'e proof-net with conclusions $\Gamma$.
\end{cor}

With the above Corollary, we are ready to establish equivalence of RL and Pomset logic.

\subsection{Equivalence of systems}
Clearly there is one-to-one correspondence between R\&B proof-structures and RL proof-structures, as there is such a correspondence on the level of links.
To show equivalence of RL and Pomset logic, it is sufficient to show that this correspondence takes proof-nets to proof-nets. And, thanks to cut-elimination, we may restrict our attention to cut-free proof-nets.

So, let $\rho$ be an RL proof-net, and let $\hat\rho$ be the corresponding R\&B proof-structure.

\begin{lem} If for some Retor\'e switching $\sigma$ of $\rho$ there is an elementary directed cycle in $\sigma$, then there is an a.e.\ circuit $\hat\phi$ in $\hat\rho$.
\end{lem}

\begin{proof}
Without loss of generality we may assume that in all $\vwpop$-links not critical for $\phi$ one of the edges is erased by $\sigma$.

We introduce the following terminology. A \emph{turn} in $\phi$ is a $\otimes$- or $\vwpop$-link, such that $\phi$ traverses both its premises and conclusion in the order: one premise --- conclusion --- the other premise; we identify such a link with a directed subpath of $\phi$.  A \emph{monotone edge} of $\phi$ is any edge of $\phi$ that is not in a turn and is not  an axiom link. A \emph{monotone path} in $\phi$ is any subpath of $\phi$ consisting of monotone edges.

Note that the only directed edges of $\phi$ occur in turns, more precisely, in its critical $\vwpop$-links. A monotone path has no directed edges, so it has two possible orientations. We call a monotone path \emph{downward} if the direction is chosen from formulas of lower complexity to formulas of higher complexity, otherwise it is \emph{upward}.

 The cycle $\phi$  has a representation as a concatenation of directed paths: \ldots--- axiom link --- downward  path --- turn --- upward path --- axiom link --- \ldots

We are going to translate subpaths of $\phi$ to  alternating paths in $\hat\phi$ satisfying the following conditions.

\begin{itemize}
\item A downward  subpath is translated to an alternating path whose first edge is regular and whose last edge is bold, and which does not have directed edges.

\item An upward  subpath is translated to an alternating path whose first edge is bold and whose last edge is regular, and which does not have directed edges.

\item An axiom link is translated to an axiom link (which is bold).

\item A turn is translated to a regular edge.

\item If the last vertex of the subpath $\alpha$ is the first vertex of the subpath $\beta$, then the same holds for their translations $\hat\alpha$ and $\hat\beta$: the last vertex of $\hat\alpha$ is the first vertex of $\hat\beta$.
\end{itemize}

\noindent
It is clear that the translated subpaths, when concatenated, produce an a.e.\ circuit in $\hat\rho$.
The translation goes as follows.

Axiom links of $\rho$ are translated to their images in $\hat\rho$ in the obvious way.

A downward  edge has one of the three possible forms
\[A-A\otimes B,\mbox{ }A-A\wpop B,\mbox{ }A-A\vwpop B.\]
 These are translated to alternating paths, respectively,
 \[A-\otimes-A\otimes B,\mbox{ }A-\wpop-A\wpop B,\mbox{ }A-\vwpop-A\vwpop B\]
  in $\hat\rho$. A downward  path is translated edge by edge. The case of an upward  path is mirror symmetric.

A turn has one of the two possible forms: \[A-A\otimes B-B\]
or
\[A-A\vwpop B-B.\]
 For the first case: the vertices $A$ and $B$ in $\hat \rho$ are adjacent to a $\otimes$-link, hence they are connected by an undirected regular edge. This gives the translation.

 For the second case, the vertices $A$ and $B$ in $\hat \rho$ are adjacent to a $\vwpop$-link, hence they are connected by a directed regular edge $A-B$. (Note that the direction of the regular edge in $\hat\rho$ in the obvious sense agrees  with the direction of the  turn in $\rho$.) This, again, gives the translation.
\end{proof}

The converse is similar, but somewhat easier, because in $\rho$ we do not have to care about alternation of edges.
\begin{lem} If in $\hat\rho$ there is an a.e.\ circuit $\phi$, then there exists  some Retor\'e switching $\sigma$ of $\rho$ with an elementary directed cycle.
\end{lem}

\begin{proof} Note that for each $\wpop$-link or $\vwpop$-link in $\hat\rho$ the  cycle $\phi$ traverses at most one of its regular edges, otherwise in $\phi$ there are two regular edges in a row. (Actually the same holds for $\otimes$-links as well, but we do not need this observation.)

We define a Retor\'e switching $\sigma$ of $\rho$ as follows.

Assume there is a $\wpop$-link in $\hat\rho$ with conclusion $A\wpop B$ traversed by $\phi$.  In the corresponding $\wpop$-link in $\rho$ erase the edge $A-A\wpop B$, if $\phi$ goes through $B-\wpop$, and erase $B-A\wpop B$, if $\phi$ goes through $A-\wpop$.

Assume there is a $\vwpop$-link in $\hat\rho$ with conclusion $A\vwpop B$ traversed by $\phi$.  In the corresponding $\vwpop$-link in $\rho$ erase the edge $A-A\vwpop B$, if $\phi$ goes through $B-\vwpop$, and erase $B-A\vwpop B$, if $\phi$ goes through $A-\vwpop$. Do not erase anything, if $\phi$ goes through $A-B$.

It is straightforward that $\phi$ translates to a directed cycle in this switching of $\rho$.
\end{proof}

This gives us the following.
\begin{cor}
Pomset logic and RL are equivalent.
\end{cor}

We conclude with one of the key results of the paper.
\begin{cor}
The sequent calculus for RL is sound and complete for Pomset logic.
\end{cor}

\begin{rem}
It is worth noting  that we could have given an alternative proof of equivalence between RL and Pomset
logic by showing directly that the sequent calculus for RL is sound and complete for Pomset logic. For that, we need to
 interpret reachability in decorated sequents as: $(X_1,\ldots,X_n)\rightrightarrows(Y_1,\ldots,Y_n)$ iff in the
 corresponding R\&B proof-net there are pairwise nonintersecting alternating paths from $X_i$ to $Y_i$, $i=1,\ldots,n$,
 starting and ending with bold edges. Then we can mimic soundness and completeness proofs for RL of this paper.

 On the other hand, we can argue that the formalism of Retor\'e proof-nets that we use is an interesting alternative to original
 R\&B proof-nets. This formalism  seems  closer to a more common  presentation of MLL proof-nets as in~\cite{Girard},
 and the correctness criterion appears as a simple generalization of the familiar Danos-Regnier criterion
 \end{rem}

\section{Conclusion and further work}
We defined a system of semicommutative linear logic, which is a non-commutative extension of multiplicative linear logic, by
means or proof-nets and by means of decorated sequent calculus. We then constructed Retor\'e logic as a degenerate variant
and showed that its equivalent to Pomset logic. As a consequence, we found an alternative presentation of Pomset logic
in terms of Retor\'e prof-nets and as a decorated sequent calculus.

The system ScMLL was discovered by the author semantically. Two non-isomorphic dual noncommutative
connectives arise when the construction of~\cite{BluteSlavnovPanangaden} for modeling Retor\'e's noncommutative
connective in probabilistic coherent spaces is generalized to a wider class of partially ordered vector spaces.
This will be shown in a forthcoming paper.

The question of  general
category-theoretic axiomatization remains open. This applies to RL (Pomset logic) as well. An attempt of categorical
axiomatization of the deep inference system BV  (conjecturally equivalent to Pomset logic) has been done
in~\cite{BluteSlavnovPanangaden}, however no kind of soundness was proven in that work.

The problem of equivalence of BV and Pomset logic itself is  a sufficiently long standing open question related to the
subject of the current paper. The author would like to believe that having now a sequent calculus, we can somehow
progress with this problem.

Finally, there is a wide field for work on extending semicommutative logic to additive and exponential fragments.
An interesting question is if there are any new modalities associated  to the new connectives. (In Pomset logic there is,
indeed, a self-dual modality for the self-dual ``before''~\cite{RetoreModality}.)

\section*{Acknowledgment}
\noindent The author is grateful to Lutz Strassburger for comments and useful discussions.

\bibliography{sslavnov bibliography}
\bibliographystyle{alpha}

\end{document}